\documentclass[11pt,a4paper]{article}
\usepackage{jheppub}
\usepackage{amsmath}
\usepackage{amssymb}
\usepackage{cancel}
\usepackage{slashed}
\usepackage{graphicx} 
\usepackage{array}

\newcommand{\bea}{\begin{eqnarray}}
\newcommand{\eea}{\end{eqnarray}}
\newcommand{\bit}{\begin{itemize}}
\newcommand{\eit}{\end{itemize}}
\def\tabspace{\vrule height 3.5ex depth 2.2ex width 0pt}
\def\half{{\frac{1}{2}}}
\def\nl{\nonumber \\}
\def\wt{\widetilde}
\def\d{\partial}
\def\n{\nabla}
\newcommand{\TO}[1] {\langle  \mathbf {T} \left \{#1\right \} \rangle}

\def\JJ {\mathcal {J}}
\def\LL {\mathcal {L}}
\def\AA {\mathcal A}
\def\OO {\mathcal O}
\def\WW {\mathcal W}

\def\TT {\mathcal T}

\def\Dm{\mathcal{D}}
\def\bDm{\bar{\mathcal{D}}}
\def\a{\alpha}
\def\da{{\dot{\alpha}}}
\def\b{\beta}
\def\bab{{\bar{ \beta}}}

\def\l{\lambda}
\def\bl{\bar{\lambda}}
\def\bi{{\bar{i}}}
\def\bj{{\bar{j}}}
\def\bk{{\bar{k}}}
\def\bbl{{\bar{l}}}

\def\s{\sigma}
\def\bs{\bar{\sigma}}

\def\p{\partial}

\def\N{\nabla}
\def\bN{\bar{\nabla}}
\def\g{\Gamma}
\def\bg{\bar{\Gamma}}
\def\li{\mathcal{L}}
\def\bli{\bar{\mathcal{L}}}
\def\le{\left(}
\def\ri{\right)}

\def\commuA{(\s_{[1} \bs_{2]} )}
\def\commuB{(\s_{[1} \Dm^\a \s_{2]})}
\def\commuC{(\Dm^\a \s_{[1} \bDm^\da \bs_{2]})}

\def\beq{\begin{equation}}
\def\eeq{\end{equation}}
\def\t{{\theta}}
\def\bt{{\bar{\theta}}}
\def\bI{{\bar{I}}}
\def\bJ{{\bar{J}}}
\def\bK{{\bar{K}}}
\def\bbL{{\bar{L}}}

\def\arr{{\rightarrow}}
\def\LRG{{LRG}}
\def\sLRG{{SLRG}}
\def\sW{{SW}}
\def\sC{{SC}}

\title{Superspace formulation of the local RG equation} 

\author[a,b]{Roberto Auzzi} 
\author[c]{and Boaz Keren-Zur}

\affiliation[a]{Dipartimento di Matematica e Fisica, Universit\`a Cattolica, \\
Via Musei 41, 25121 Brescia, Italy}
\affiliation[b]{ INFN Sezione di Perugia, \\ Via A. Pascoli, 06123 Perugia, Italy}
\affiliation[c]{Institut de Th\'eorie des Ph\'enom\`enes Physiques, EPFL,\\
CH-1015 Lausanne, Switzerland}

\emailAdd{roberto.auzzi@unicatt.it}
\emailAdd{boaz.kerenzur@epfl.ch}

\abstract{
We present the superspace formulation of the local RG equation, 
a framework for the study of supersymmetric RG flows in which the constraints of holomorphy and $R$-symmetry are manifest.
We derive the consistency conditions associated with super-Weyl symmetry off-criticality
 and initiate the study of their implications.
As examples, we derive an expression for the $a$-function,
and present an analog of the $a$-maximization equation, which is valid off-criticality.
We also apply this machinery to the study of conformal manifolds and give a simple proof that the metric on such manifolds is K\"ahler. 
}


\begin{document}

\maketitle


 \section{Introduction}
\label{sec_intro}

Renormalization group (RG) flows describe a trajectory in the space of theories, induced by a change of scale. 
It is a major challenge of quantum field theory to characterize these trajectories, and understand their structure.
One of the theoretical tools used for this purpose is the local RG equation, formulated in 
\cite {Osborn:1989td,Jack:1990eb,Osborn:1991gm} 
(for recent reviews, see also \cite{Nakayama:2013is,Baume:2014rla,Jack:2013sha}). 
The idea is to consider the RG evolution as a dilatation symmetry transformation, which is explicitly broken by interaction terms in the Lagrangian. 
Promoting the coupling constants $\l$ to background fields $\lambda(x)$, the symmetry can be formally restored by assigning them with compensating transformation properties, determined by their $\beta$-function and anomalous dimensions. Moreover, introducing a background metric $g_{\mu\nu}$, the global dilatation symmetry can be promoted to a local Weyl symmetry.
The local RG (\LRG) equation is nothing but the anomalous Ward identity for the generalized Weyl symmetry. 
Roughly speaking, it takes the following form 
\bea
\Delta^{W}_\s \WW [g,\l]\equiv   \int d^4x \s \le 2 g_{\mu\nu}\frac{\delta}{\delta g_{\mu\nu}(x)}
 +\beta \cdot \frac{\delta}{\delta \lambda(x)}+\ldots\ri \WW [g,\l]=\AA^W_\s[g,\l]~,
\eea
where $\WW$ is the generating functional for correlation functions of composite operators, $\Delta_\s^W$ is a generator for the Weyl symmetry transformation,  and $\AA_\s^W$ is the anomaly function, which is a local function of the background fields. These objects will be defined in more detail below.
 The major results of this formalism are based on the  Wess-Zumino consistency conditions for this anomaly 
 \bea
 \label{eq_Weyl_Wess_Zumino}
 \left [\Delta^W_{\s_2} , \Delta^W_{\s_1} \right] \WW = \Delta^W_{\s_2}\AA_{\s_1}- \Delta^W_{\s_1}\AA_{\s_2}&=&0~.
 \eea 
 This equation gives non-trivial relations between the various anomaly coefficients, which are functions of the coupling constants. 
 These constraints on functions of the couplings can be translated into constraints on the RG flow.
A prominent example is the proof for the irreversibility of perturbative unitary RG flows \cite {Jack:1990eb,Osborn:1991gm}.

The goal of this work is to apply the formalism of the local RG equation to the study of supersymmetric RG flows.
Supersymmetric RG flows are known to have a rich structure and non-trivial properties, such as the non-renormalization theorems for the superpotential  \cite{Seiberg:1993vc} and the exact formula for the $\beta$-function in gauge theories
\cite{Novikov:1983uc,Shifman:1986zi}.  
The derivation of these results is based on two properties of supersymmetric theories -- $R$-symmetry and holomorphy.
Supersymmetry can be introduced to the formalism of the \LRG~equation by directly specializing the field content and the couplings to the supersymmetric case, leading to some interesting constraints \cite{Freedman:1998rd} (and more recently \cite{Jack:2013sha} and \cite{Jack:2014pua}). 
However, the formulation of the equation in components does not exploit the power of holomorphy. 
For this purpose we formulate the \LRG~equation using
 superspace notation\footnote{Superfield formalism simplifies also the derivation of renormalization group
equation for SUSY breaking parameters, using analytic continuation
into superspace \cite{ArkaniHamed:1998kj}.}, with a supergravity background. 
The new equation we find, the superspace local RG (\sLRG) equation, corresponds to the Ward-identity of the super-Weyl symmetry off-criticality, 
and it makes holomorphy and the relation with $R$-symmetry manifest.

In the framework described here, the non-renormalization theorem and the NSVZ formula \cite{Novikov:1983uc,Shifman:1986zi}
 appear as a natural part of the construction. In addition, we find a generalization of the consistency conditions \eqref{eq_Weyl_Wess_Zumino}, and begin the exploration of their implications. As in the framework of the LRG equation, we define a function $\tilde a$, which is a continuation off-criticality of the $a$ coefficient in the Weyl anomaly, and find a relation between its derivative with respect to the coupling $\lambda^I$ and the $\beta$ function:
\bea
\label{eq_intro_almost_gradient_flow}
\frac{\d}{\d\l^I}\tilde a&=& \bar \beta^\bJ\chi_{I\bJ}-\beta^J (\d_J\bar w_I - \d_I \bar w_J)
\eea
where $\chi$ and $\bar w$ are functions of the couplings appearing as anomaly coefficients. 
A similar result is derived from the \LRG~consistency conditions.
A remarkable result which appears only in the \sLRG~framework is that the tensor $\chi_{I\bJ}$ can be written as a gradient of a function we denote by $\Omega_I$, 
plus a function of other anomaly coefficients ($w_\bJ, \xi^{1}_{[IK] \bj},\zeta^1_{A I}$):
\beq
\chi_{I \bJ}- \p_I w_\bJ 
-   2i \xi^{1}_{[IK] \bJ}  \b^K  
  +  \zeta^1_{A I} \p_\bJ \g^A
=  - \p_\bJ \Omega_I \, .
 \label{eq_omegone2_intro}
\eeq
In addition, we find a consistency condition which 
coincides with the $a$-maximization formula \cite{Intriligator:2003jj}  at the fixed point. 
Plugging these results back into \eqref{eq_intro_almost_gradient_flow}, 
we find the following expression for $\tilde a$ in terms of the anomalous dimension matrices $\gamma$
\bea
 \tilde  a
&=&-\frac{1}{128 \pi^2} \text{Tr} [\gamma^2] + \frac 1 {192\pi^2} \text{Tr}[\gamma^3 ]  -\beta^I \wt  \Omega_I +\text{const} \, ,
\eea
where $ \tilde \Omega_I=\Omega_I-\bar w_I $. 
A similar ansatz for $\tilde{a}$,
where $ \tilde \Omega_I$ was an undetermined function,
 was conjectured in \cite{Freedman:1998rd} and checked to be consistent up to 4-loop order  
(see e.g.~\cite{Kutasov:2004xu,Barnes:2004jj,Jack:2013sha,Jack:2014pua}),
 under some assumptions regarding the anomalous dimensions, on which we comment below.
 In \cite{Kutasov:2003ux,Kutasov:2004xu,Barnes:2004jj} 
 $\wt\Omega_I$ is interpreted as a Lagrange multiplier imposing the vanishing of the $\beta$ function at the fixed point.

Another potential application of the \sLRG~framework is the study of manifolds of fixed points (conformal manifolds)\cite{Leigh:1995ep}. To demonstrate the usefulness of this approach we give a simple proof for the fact that the Zamolodchikov metric on this manifold is 
K\"ahler (for an earlier proof see also \cite{Asnin:2009xx}). We believe that the same machinery is suitable for the investigation of additional properties of the conformal manifold (as in \cite{Kol:2002zt} and \cite{Green:2010da}).

The paper is organized as follows: In section \ref{sec_symmetry} we define the basic ingredients used in our formalism, namely the background fields and the generating functional $\WW$, and present the generalized super-Weyl (\sW) symmetry. 
We highlight the differences between the general formulation of the \LRG~equation and the superspace formulation (\sLRG), which are the consequences of the choice of the holomorphic scheme. 
Using the generator of the \sW~symmetry we define a generator of infinitesimal super-conformal (\sC) symmetry transformations acting on the sources. Using these expression we easily extract some constraints on the RG flow related to $R$-symmetry.
We conclude with a discussion of the equation in components, and some technical aspects regarding the \sW~variation of functions.

In section \ref{sec_anomaly} we define the super-Weyl anomaly, and its consistency conditions. There are several aspects in which this discussion differs from the presentation of the consistency conditions in \cite{Osborn:1991gm}. 
First, following the methodology of \cite{Keren-Zur:2014sva}, we introduce non-gauge-invariant \sW~anomaly terms which are necessary in the presence of chiral anomalies.  Next, we discuss a peculiar feature of the Wess-Zumino consistency conditions of the \sW~anomaly, which implies that  some of the consistency conditions are given only up to some unknown functions.
 Finally, and most importantly, we find many more equation than in the \LRG~framework.
This is to be expected, as the \sLRG~equation describes a larger symmetry. 
An initial exploration of their implications, as described above, is given in section \ref{sec_implications}.
Some of the relevant definitions and formulas appear in the appendix. Throughout this paper we use the notations of \cite{Buchbinder:1995uq}.

 \section{The generalized super-Weyl symmetry}

\label{sec_symmetry}

 \subsection{Background sources and the generating functional}
   \label{sec_bg_sources}
Our goal is to study supersymmetric RG flows in the vicinity of superconformal fixed points.
The main tools at our disposal are a set of background source fields $\JJ$ and a functional of these fields $\WW[\mathcal J]$, which generates the renormalized correlation functions of the composite operators in the superconformal theory
\bea
\label{eq_W}
 \frac{\delta}{\delta \JJ_1(z_1)}\ldots \frac{\delta}{\delta \JJ_n(z_n)} \WW [\mathcal J] \Big|_{\JJ=0}&=&i^{1-n}\TO{\OO_1(z_1)\ldots \OO_n (z_n)}~.
\eea 
Such a functional is schematically given (if the fixed point has a Lagrangian description) by
\bea
\label{eq_Lagrangian_W}
e^{-i\WW[\JJ]}&=&\int D \Phi ~e^{-i\int \left (\LL_0[\Phi] + \JJ \OO + \LL_{c.t.}(\JJ,\Phi)\right)}
\eea
where $\Phi$ are the dynamical fields and $\LL_0[\Phi]$ is the Lagrangian at the fixed point.
 A non-dynamical supergravity background,
 which is a useful way to realize the super-Weyl symmetry of the theory,
  is implicit in this notation. 
In order to ensure the finiteness of the correlation functions, $\WW$ must contain a set of local counterterms $\LL_{c.t.}(\JJ,\Phi)$, which consists of functions of the sources and their derivatives. The existence of these counterterms is responsible for the appearance of the anomaly.
It should be emphasized, however, that eq. {\eqref{eq_W}} is sufficient as a definition for $\WW$, and that the formalism is valid also for conformal fixed points with no Lagrangian description.

The functional derivatives in eq. \eqref{eq_W} evaluated with a small, non-zero, background value for $\JJ$ gives the correlation functions in a theory where a deformation $\int \JJ\OO$ is turned on. In that sense, the background source fields $\JJ$ can be understood as the coupling constants promoted to a coordinate dependent field. There is yet another way in which we can take advantage of these background fields: in a given theory, some of the global symmetries of the fixed point may be broken explicitly by the deformations. However,  one can formally restore these symmetries by assigning transformation properties to the background fields, which compensate for the non-invariance of the theory. The background fields $\JJ$ thus play several different roles in our framework: coupling constants, sources for composite operators, and compensators for broken symmetries.

The idea of compensating background fields is well known: for example, it has been exploited in the past for the study of broken flavor symmetries \cite{Gasser:1983yg}. The application of this idea to Weyl symmetry, which was formulated in \cite{Osborn:1991gm},
is the relevant realization of this idea for our discussion. In that paper, the background fields were used to define a generalized form of the Weyl symmetry, which is valid off-criticality, and the consistency conditions associated with this symmetry \cite{Bonora:1983ff}
 lead to non-trivial constraints on RG flows. Here we specialize to supersymmetric theories and we take this idea one step further: we introduce superspace notations, and consider the sources as compensators of the super-Weyl (\sW) symmetry of the super-conformal (\sC) fixed point. 

We will restrict our discussion to RG flows induced by marginal deformation. 
As discussed in \cite{Green:2010da}, in a flat background there are only two possible options
for such a deformation: chiral operators integrated over $d^2 \t$ 
(superpotential deformations) which we denote by $\OO_I$, 
or generic operators integrated in $d^4 \t $
(K\"ahler deformations) which we denote by $J_{\bi j}$ 
(the structure of the indices will be explained in the next section). 
In order for the K\"ahler deformation $J$ to be  marginal,
it should correspond to a conserved current ($\bDm^2 J=0$), 
and then $\int d^4 \t J$ does not deform  the Lagrangian.
If the current $J$  is not conserved, then the deformation is irrelevant 
by unitarity and will be neglected in the following analysis.

A general discussion of RG flows requires introduction of an infinite set of sources, for an infinite set of operators. However, for the discussion of RG flows induced by marginal deformations, with no mass parameters,  it is sufficient to introduce sources for the marginal operators only. 
We will consider a chiral source $\lambda^I$ and a real source $Z^{j \bi}$ defined such that\footnote{We use the conventions 
$\frac{\delta \lambda(z_+)}{\delta \lambda(z'_+)}=\varphi^{-3}\delta^{(4)}(x-x')\delta^{(2)}(\theta-\theta'),~$
$\frac{\delta Z(z)}{\delta Z(z')}=E\delta^{(4)}(x-x')\delta^{(2)}(\theta-\theta')\delta^{(2)}(\bar \theta-\bar\theta')$ 
}
\bea
\frac{\delta}{\delta\lambda^I(z)}\WW \equiv[\OO_I(z)]
\qquad
\frac{\delta}{\delta Z^{j \bi}(z)}\WW \equiv [J_{\bi j}(z)]~.
\eea
In our notations, the brackets mean that the operator is renormalized.
The source $Z$ can be understood as the normalization of the kinetic term.
For example, in the case of a Wess-Zumino model the coupling of the sources takes the form
\bea
\int d^8z E^{-1}Z^{j \bi} J_{\bi j} +\int d^6z \varphi^3 \l^I \OO_I   + {\rm c. c.}
\eea
where
\bea
J_{\bi j} \equiv  \bar{\Phi}_\bi \Phi_j \, , \qquad
\l^I \OO_I  \equiv  \l^{ijk} \Phi_{i} \Phi_j \Phi_k  \, .
\eea
Alternatively, one can work in components, set the normalization of the kinetic terms to unity and absorb the wavefunction renormalization into the bare fields.
However, by doing that one loses information about the constraints imposed by holomorphy. Our choice to work in the holomorphic scheme forces us to introduce $Z$ as an independent source, and to work with the non-physical holomorphic coupling constants $\lambda$.

As mentioned above, the generating functional is defined in a curved supergravity background. We will use the old-minimal formulation of supergravity, and the notations of \cite{Buchbinder:1995uq}. For the purpose of the discussion here, it is sufficient to mention that the supergravity fields include a superfield containing the background metric $H^{\a\da}$ and a chiral field $\varphi$ known as the chiral compensator.
$\varphi$ will play an important role in the realization of the super-Weyl symmetry on $\WW$.
These fields can also be understood as the sources for the supercurrent
 (Ferrara-Zumino multiplet \cite{Ferrara:1974pz}) $\mathcal T_{\a\da}$ and  
a chiral operator $\mathcal T$
\bea
\label{eq_supercurrent}
\frac{\delta}{\delta H^{\a\da}(z)}\WW = -\frac{1}{2} [\mathcal T_{\a\da}(z)] \, ,
\qquad 
\frac{\delta}{\delta \varphi ^3(z)}\WW \equiv \frac 1 3 [ \mathcal T(z)]~,
\eea
with which we can write the supersymmetric generalization of the conservation
of energy momentum tensor:
\bea
0&=&
\bar{\mathcal D}^\da[T_{\a\da}] + \frac{2}{3}\mathcal D_\a[ T]~.
\eea
Using this Ward identity we can write the $\theta^2$ component of the chiral superfield $\mathcal T$ as
\bea
\label{eq_T_component}
\mathcal T |_{\theta^2} = \frac{1}{2}  T^\mu_\mu+ i\frac 3 4 \d_\mu j^\mu_5
\eea
where $T_{\mu\nu}$ is the energy momentum tensor and $j^\mu_5$ is the lowest component of the supercurrent, 
which at the fixed point coincides with the $R$ current.

 \subsection{Flavor symmetries}
  \label{sec_flavor_symmetry}
As mentioned in the previous section, the coupling constants which induce the RG flow break explicitly some of the symmetries of the fixed point. 
In this section we discuss global, internal, perhaps anomalous, symmetries, which we will denote by $G$ 
and refer to as "flavor" symmetries. We will assume that the marginal operators $\OO_I$ reside in some representation of the symmetry group, and their transformation rule is given by
\bea
\delta_\omega^G \OO_I &=& -\omega^A \, \OO_J (T_A)^J_I
\eea
where $(T_A)^I_J$ are the generators of the symmetry, and $\omega^A$ is some constant transformation parameter. Obviously, the global symmetry can be restored (up to anomalies) by assigning the following transformation properties to the fields
\bea
\delta_\omega^G \l^I&=& \omega^A \, (T_A)^I_J \l^J~.
\eea

The framework of the background sources and generating functional $\WW$ can be used to generate the Ward identities for the flavor symmetries. This is done by promoting the global symmetry to a local one and using background gauge fields. Conveniently, the necessary background fields were already introduced in the above discussion. Indeed, the sources $Z$ can be interpreted either as wavefunction renormalization, or equivalently, as gauge fields which act as sources for the Noether currents of the theory. 
In the example of a Wess-Zumino model,
we can use the more familiar notation $Z^{i\bj}\equiv (e^{-V})^{i\bj}$, where $V$ is a real vector superfield\footnote{In our normalization the $\theta \sigma^\mu \bar\theta$ component of $V$ is twice the ``canonical'' gauge field $A_\mu$,
which is defined as the one appearing in the covariant derivatived of fermions as
$D_\mu \psi_\Phi = \partial_\mu \psi_\Phi -i A^A_\mu T_A \psi_\Phi$. \label{footnot}}, and 
we can write
\bea
Z^{i\bj}\Phi_i\bar \Phi_{\bj}\equiv \Phi_i (e^{-V})^{i\bj}\bar \Phi_{\bj}
\eea
and Noether currents can be defined as
\bea
(e^V)_{\bj i}(T_A )^i_k\frac{\delta}{\delta (e^V)_{\bj k}}\WW&=&
-[\Phi_i (T_A)^i_k  (e^{-V})^{k \bar{j} } \bar{\Phi}_{\bar{j}}] \equiv [J_A] \, .
\eea 
The indices $i$ ($\bar{i}$) run in the fundamental (antifundamental)
of the global symmetry group (which may be broken by the chiral couplings $\l^I$).
It is always possible to decompose the representation of the chiral couplings
$\l^I$ in product of fundamentals; from this decomposition we can 
infer how $e^V$ acts on the couplings $\l^I$.

A few words about our index convention are in order.
A super-gauge transformation can be parameterized by the exponential of a chiral gauge parameter: $(e^{i \Lambda})^i_j$,
 where $\Lambda=\Lambda^A T_A$ and  $T_A$ is a basis of hermitian matrices.
As a mnemonic device, the indices for the sources are chosen as:
\beq
\l^I \, , \qquad \bl^{\bI} \, , \qquad
(e^V)_{\bj k} \, , \qquad (e^{-V})^{i \bj} \, . 
\eeq
The gauge field $e^V$ and its inverse transform as:
\beq
(e^V)_{\bj k}     \rightarrow   (e^{i \bar{\Lambda}  })_\bj^\bbl  (e^V)_{\bbl i}   (e^{-i \Lambda})_k^i  \, ,
\qquad
(e^{-V})^{i \bj}     \rightarrow   (e^{i \Lambda  })_k^i  (e^{-V} )^{k \bbl}   (e^{-i \bar{\Lambda}})_\bbl^\bj \, . 
\eeq
The chiral coupling transform as:
\beq
\lambda^I \rightarrow (e^{i \Lambda})^I_J \lambda^J \, , \qquad
\bar{\lambda}^{\bar{I}} \rightarrow \bar{\lambda}^\bJ (e^{-i \bar{\Lambda}})_\bJ^{\bar{I}} \, .
\eeq
The representation according to which $\l^I$ transforms can be obtained from some opportune tensor product of the fundamental
representation, which has indices $i$.
It is sometimes convenient to write objects as $Y^i_j$ and $\bar{Y}^\bi_\bj$
in term of adjoint indices: $Y^i_j= Y^A (T_A)^i_j $, 
where matrices $(T_A)^i_j$ are the hermitian generators of the fundamental representation.
We can then write the same object in an arbitrary representation;
for example, in the representation of the chiral coupling: $Y^I_J= Y^A (T_A)^I_J $.
 We denote with  $\bar{T}_A$ the complex conjugate
 of the generator $T_A$: $(\bar{T}_A)^\bi_\bj= ((T_A)^i_j)^*=(T_A)^j_i$;
 with this notation $\bar{Y}^\bi_\bj= \bar{Y}^A (\bar{T}_A)^\bi_\bj $.
The generators of the anti-fundamental representation are:
 $(\tilde{T}_A)^\bi_\bj=-(\bar{T}_A)^\bi_\bj$.

In general, one may consider chiral operators, and associated chiral sources, which are invariant under all flavor symmetries of the fixed point.
However, the occurrence of such singlet chiral sources is rare (in fact the only example we are aware of are the holomorphic gauge couplings in pure super Yang-Mills theory). Moreover, in most cases it is impossible to write a holomorphic function of the $\lambda$'s which is a singlet of the symmetries. 
We refer to this property as the "absence of chiral singlets".
In theories where it applies, we will show below that it can be used to derive strong constraints on the flow.

We are now ready to define a generator for the flavor symmetry transformations
\bea
\label{eq_Delta_G}
\Delta_\Lambda^G &\equiv& 
\int d^6z\varphi^3~ i(\Lambda\l)^I\frac{\delta}{\delta \lambda^I} + c.c\nl
&&-\int d^8z  E^{-1}(i (e^V \Lambda)_{\bi j} -i( \bar \Lambda e^V)_{\bi j})\frac{\delta}{\delta (e^V)_{\bi j}}~,
\eea
The anomalous invariance equation for $\WW$ is written as 
\bea
\Delta_\Lambda^G ~\WW=
\AA_\Lambda^G~,
\eea
where $\AA_\Lambda^G$ is the anomaly involving background gauge fields, which we will further discuss in section \ref{sec_chiral_anomalies}. Anomalies involving dynamical gauge fields are realized in this formula by assigning transformation properties to the holomorphic gauge couplings (see section \ref{sec_NSVZ}).
Taking the transformation parameter $\Lambda(z)$ to be the chiral delta function $\delta_+(z-y)$ and evaluating in the flat and constant background, we find the operator equation
\bea
\label{eq_flavor_operator_WI}
(T_A\l)^I [\OO_I(y)]&=&-\frac{1}{4}\bar D^2 [J_A (y)] ~.
\eea 
In this equation the anomaly vanishes because we take the background sources to be constant. Finally, the Ward identity for a correlation function is given by
\bea
\label{eq_WI_correlation_function}
\Delta_\Lambda^G
\frac{\delta}{\delta \lambda^{I_1}(z_1)}\ldots \frac{\delta}{\delta \lambda^{I_n}(z_n)}\WW 
&=&
\left [\Delta_\Lambda^G, \frac{\delta}{\delta \lambda^{I_1}(z_1)}\ldots \frac{\delta}{\delta \lambda^{I_n}(z_n)}\right] \WW \nl
&&+\frac{\delta}{\delta \lambda^{I_1}(z_1)}\ldots \frac{\delta}{\delta \lambda^{I_n}(z_n)}\AA_\Lambda^G~.
\eea

 \subsection{The super-Weyl symmetry}
  \label{sec_Weyl_symmetry}

In a supergravity background,  $\WW$ possess another important symmetry, which is broken by the marginal deformations -- the super-Weyl (SW) symmetry.  This symmetry is realized as a variation of the chiral compensator  $\delta_\s^\sW\varphi=\s\varphi$. 
At the fixed point this symmetry is generated by
\bea
\Delta_\s^\sW &\equiv & 
\int d^6z~\s\varphi\frac{\delta}{\delta \varphi} + c.c
\eea

As discussed in \cite{Osborn:1991gm},  in order to formally restore the Weyl symmetry broken by quantum effects,
the background sources $\lambda$ and $Z$ (or equivalently $V$) must be given Weyl transformation properties, which compensate for the running of the couplings. The Weyl symmetry can thus be generated by an operator similar to the one defined in eq. \eqref{eq_Delta_G}:
\bea
\Delta_\s^\sW &\equiv & 
\int d^6z~\varphi^3\s\left (3\frac{\delta}{\delta \varphi^3} +   b^I \frac{\delta}{\delta \lambda^I}\right ) + c.c\nl
&&+
\int d^8z E^{-1}\big( \sigma \Gamma^A (e^V T_A)_{\bi j} +\bs  \bar \Gamma^A( \bar T_A e^V)_{\bi j}\big)\frac{\delta}{\delta (e^V)_{\bi j}}~.
\eea
We rewrote the variation of $\varphi$ to allow usage of eq. \eqref{eq_supercurrent}.
$b^I=b^I(\l)$ is a holomorphic function of chiral couplings we will refer to as the holomorphic $\beta$-function, 
and $\Gamma^A=\Gamma^A(\l,\bl,e^V)$ is a superfield which contains the anomalous dimensions of the dynamical fields. In the following sections we will discuss these functions and the model independent constraints imposed on them by the different symmetries of the theory.

As mentioned above, a full analysis of the Weyl symmetry off-criticality requires introduction of sources for each of the composite operators in the spectrum, and not only the marginal ones. However, for the discussion of a flow in which none of the dimensionful parameters obtains a non-zero VEV, it suffices to introduce only dimensionless sources. There is one exception we are aware of -- there could be contributions to the local RG equations from D-terms involving chiral dimension 2 operators (e.g. $\int d^8z E^{-1} \eta \Phi^2$, where $\eta$ is some function of the couplings). 
However, such terms can be eliminated by a choice of improvement (see, e.g.  discussion 
in \cite{Komargodski:2010rb,Nakayama:2012nd}),
 and we leave the study of such terms for the future.

The supersymmetric local RG (\sLRG) equation, which is the subject of our discussion, is nothing but the (anomalous) Ward identity for this symmetry
\bea
\Delta_\s^\sW \WW &=&
\AA_\s^\sW~.
\eea
As an operator equation (going to the flat and constant background) this can be written as
\bea
\label{eq_operator_equation}
[\TT]&=&-\le b^I [\OO_I] -\frac 1 4 \Gamma^A \bN^2 [J_A]\ri
\eea
Using the operator equation \eqref{eq_flavor_operator_WI} one can rewrite this in a more familiar form
\bea
\label{eq_operator_equation_beta}
[\TT]&=&-\beta^I [\OO_I]
\eea
where we identify the physical $\beta$-function 
(which corresponds to the numerator of NSVZ \cite{Novikov:1983uc}) as
\bea
\label{eq_def_beta}
\beta^I&\equiv& b^I + \Gamma^A(T_A\l)^I ~.
\eea
Plugging this into eq. \eqref{eq_T_component}, and taking the real part, we find 
the expected trace anomaly
\bea
T_\mu^\mu= \beta^I [O_I]+c.c
\eea
where $O_I$ is minus the $\theta^2$ component of $\OO_I$.

\subsection{The superconformal symmetry}
\label{sec_superconformal}
The superconformal symmetry can be defined as a combination of super-diffeomorphisms and super-Weyl transformations which keep the superspace interval, and equivalently, the chiral compensator $\varphi$, fixed.
Following \cite{Osborn:1998qu,Erdmenger:1998tu}, the infinitesimal diff transformations, corresponding to superconformal symmetry are parameterized by a vector $h^\mu$ and a spinor $\eta^\a$ (which is a function of $h$) and generated by the operator
\bea
\LL^{diff}_{h}&\equiv&
 h^{\mu}\d_{\mu} +\eta^\a \d_\a +\bar \eta_\da \bar \d^\da~.
\eea
Under such diff-transformation of  the coordinates the compensator $\varphi$ transforms as
\bea
\delta_h\varphi&=&\s_h \varphi
\eea
where $\sigma_h$ is some chiral superfield which is a function of the transformation parameters.
 This variation can be canceled by a further Weyl rescaling, generated using the operator $\Delta_\s^\sW$. Combining the two operations, we can define a generator of superconformal transformations which is schematically given by:
\bea
\label{eq_SC_generator}
\Delta^{\sC}_h&=&
\int d^8z E^{-1} 
\LL^{diff}_{h} [\JJ]\frac{\delta}{\delta \JJ}
- \Delta^\sW_{\sigma_h}~.
\eea
 The transformation properties of the composite operators under the SC symmetry are obtained by computing the commutator
\bea
\label{eq_commutator_for_conformal}
\delta^{\sC} \left (\frac{\delta}{\delta \JJ}\WW\right)&\equiv&
\left [ \Delta^{SC}_{h},\frac{\delta}{\delta \JJ}\right]\WW~.
\eea

The super-diff parameters $h$ and $\eta$ corresponding to infinitesimal dilatations (with a scale factor $d$)  and $R$ symmetry rotations (with a phase $r$) are 
\bea
h^\mu&=&
x^\mu d+2(\theta\s^\mu\bar\theta)r\nl
\eta^\a&=& \theta^\a\le \frac 1 2 d+ir\ri~.
\eea
The corresponding infinitesimal diff generator and Weyl transformation parameters are
\bea
\label{eq_s_d_r}
\LL^{diff}_{h(d,r)} &=&
d (x^\mu\d_\mu+\frac 1 2 \theta^\a \frac{\d}{\d \theta^\a} + \frac 1 2 \bar \theta^\da\frac{\d}{\d\bar \theta^\da})
+ir ( \theta^\a \frac{\d}{\d \theta^\a} -  \bar \theta^\da\frac{\d}{\d\bar \theta^\da})\nl
\s_{h(d,r)}&=&d-i\frac 2 3 r~.
\eea

\subsubsection{$R$-symmetry and non-renormalization theorems}
\label{sec_R_constraints}
Let us write the generator of $R$ symmetry off-criticality
using eqs. \eqref{eq_SC_generator}
\bea
\label{eq_R_generator}
\Delta^{\sC}_{h(r)}&=&
 \int d^6z~\varphi^3\left ( (\LL^{diff}_{h(r)}\l)^I + i\frac {2} 3r  ~b^I \frac{\delta}{\delta \lambda^I}\right ) + c.c\nl
&&+ \int d^8z~E^{-1}  \left ( (\LL^{diff}_{h(r)} e^V)_{i\bj} 
+i\frac 2 3 r\left ( (e^V \Gamma)_{\bi j}-( \bar \Gamma e^V)_{\bi j} \right)\right)
\frac{\delta}{\delta (e^V)_{\bi j}}+ c.c~.
\eea
The first implication of this expression is that if we want to avoid assigning a non-vanishing, coupling dependent, $R$-charge to the background gauge fields, we must consider only hermitean $\Gamma$ matrices
\bea
\label{eq_non_renormalization}
\bar{\Gamma}^\bk_\bi (e^V)_{\bk j}&=& (e^V)_{\bi k}    \Gamma^k_j ~.
\eea
This constraint will be applied in all the results described in this paper.

Another constraint, with far reaching implications, is derived from the requirement that the $R$-symmetry acts linearly on the sources.
By inspection of \eqref{eq_R_generator}, this can be satisfied only if the holomorphic $\beta$-function can be written as
\bea
\label{eq_b_and_q}
b^I&=& (q\lambda)^I
\eea
where $q^I_J$ is some matrix of numbers which is at the moment unconstrained. We conclude that the holomorphic $\beta$ function, 
as defined here, is either vanishing or linear in the couplings. 
This result is consistent with the non-renormalization theorem for the superpotential   \cite{Seiberg:1993vc}. 
Indeed, superpotentials terms have vanishing holomorphic $\beta$-function\footnote{Due to an inherent ambiguity in our definition of $b$ , see section \ref{sec_ambiguity},
 there could be parameterization of the local RG equation 
 in which the holomorphic beta function $b$ is linear in the coupling. 
 This has no physical significance, because the physical beta function $\beta$ is independent of this ambiguity.}, and 
as explained in the next section, this corresponds to the fact that they have $R$-charge 2. 

Our conclusion seems to be in contradiction with the well known fact that the holomorphic $\beta$-function of the holomorphic gauge couplings $S\equiv \frac{4 \pi}{g_h^2} - i \frac{\Theta}{2 \pi}$, is a constant. 
However, as a source for the marginal deformation associated with a gauge theory, we use the background field $\lambda_G=e^{-S}$
(see appendix \ref{sec_NSVZ} for a more detailed descriptions of our conventions for gauge couplings).
This basis for the background sources has the virtues of transforming linearly under the Konishi symmetry, and vanishing in the limit of the free theory, and therefore it is more suitable to our formalism. In this basis, we find that holomorphic $\beta$-function is linear, in agreement with our discussion above.

\subsubsection{Anomalous super-Weyl weights}

Using the commutator with $\Delta^{SC}_{h(d,r)}$ we can compute the variation of the renormalized operators. 
We find the following transformation properties, evaluated with flat background:
\bea
\label{eq_r_and_d_transformations_of_operators}
\delta^{\sC}_{d} [\OO_I]&=&
-\LL^{diff}_{h(d)}[\OO_I]
-d \left (3\delta_I^J + \gamma^J_I+ \wt \gamma^J_I-\Gamma^A(T_A)_I^J\right)[\OO_J]\nl
\delta^{\sC}_{r} [\OO_I]&=&
-\LL^{diff}_{h(r)}[\OO_I]
+ i r\left (2\delta_I^J -\frac 2 3 q_I^J \right) [\OO_J]\nl
\delta^{\sC}_{d} [J_A]&=&
-\LL^{diff}_{h(d)}[J_A]-d \left (2\delta_A^B +2\gamma^B_A\right) [J_B]\nl
\delta^{\sC}_{r} [J_A]&=& 
-\LL^{diff}_{h(r)}[J_A]
\eea
where the generator of superdiffs is defined in \eqref{eq_s_d_r} and the anomalous dimension matrices are given by
\bea
\label{eq_anomalous_dimension}
\gamma_I^J=\d_I\beta^J
\qquad
\wt \gamma_I^J=\d_I  \Gamma^A  (T_A\l)^I
\qquad
\gamma_A^B=\d_I \Gamma^B (T_A \l)^I~.
\eea
Basing on the second line of \eqref{eq_r_and_d_transformations_of_operators} we can also define the $R$ charge matrix $T_R$ for the composite operators 
\bea
\label{eq_R_charge}
(T_R)_I^J\equiv2\delta_I^J -\frac 2 3 q_I^J
\eea
where $q$ is the matrix defined in \eqref{eq_b_and_q}. At the fixed point, where $\Gamma^A(T_A\l)=-(q\l)^I$, we find the expected relation between the anomalous dimension and the $R$ charge
\bea
\Gamma^AT_A&=&\frac 3 2 (T_R - 2)~.
\eea

\subsubsection{Primary operators}
It is also useful to compute the action of the special superconformal symmetry on the operators
(with the corresponding fermionic transformation parameters $s^\a$). At the fixed point, one would expect the primary operators $\OO_I$ to be annihilated by the generator of this transformations. The corresponding \sW~transformation parameter is given by
\bea
\s_{h(s)}&\propto& 
s^\a \theta_\a~.
\eea
Using the commutator \eqref{eq_commutator_for_conformal} we find the following result 
\bea
\delta^{SC}_{s} [\OO_I]&\propto&\bar s_\da \d_I\bar \Gamma^A\bN^\da [J_B]~.
\eea
We conclude that at the fixed point
\bea
\d_I\bar \Gamma^A=0 \qquad \forall ~\OO_I ~\text{primary}~.
\label{priprimary}
\eea
Due to the Ward identities of the flavor symmetries, there are linear combinations of the marginal operators, $(T_A\l)^I\OO_I$, which can be written as derivatives of currents, and therefore are not primary. 
Such operators do not necessarily satisfy condition \eqref{priprimary}.

 \subsection{Ambiguity}
 \label{sec_ambiguity}

In the presence of a non-anomalous $U(1)$ symmetries, there is an equivalence class of local RG equations, 
related by addition of $U(1)$ Ward identities
\bea
\Delta_\s^\sW \to (\Delta_\s^\sW )'=\Delta_\sigma^\sW +\Delta_{\Lambda=ia\sigma}^G~,
\eea
where $a^A$ is a vector of numbers\footnote{In principle we might contemplate the possibility
that $a^A$ as is a non-constant holomorphic function of the chiral couplings,
 but this would lead to a contradiction with the argument from 
 the previous section regarding the linear transformation of the sources under the $R$-symmetry.
}
defining the linear combination of $U(1)$ symmetries added to the local RG equation (in fact, as will be explained in sec. \ref{sec_chiral_anomalies}, this ambiguity exists also in the case of anomalous $U(1)$s).
This ambiguity can be translated into an ambiguity in the holomorphic function $b$ and the anomalous dimension $\Gamma$
\bea
\Delta_\s^\sW(b^I,\Gamma^A) \to \Delta_\sigma^\sW ({b^I}',{\Gamma^A}') =\Delta_\s^\sW(b^I,\Gamma^A)+\Delta_{\Lambda=ia\sigma}^G~,
\label{ambi1}
\eea
where
\bea
{b^I}'&=&b^I - a^A(T_A\l)^I\nl
{\Gamma^A}'&=&\Gamma^A + a^A~.
\label{ambi2}
\eea
Under this transformation the $R$ charge defined in \eqref{eq_R_charge} changes as
\bea
T_R'&=&T_R +\frac 2 3 a^AT_A~.
\eea

Recall that in the local RG
equation we are keeping only the information about
composite operators, while the elementary fields, which are not assigned with independent sources, can be left unspecified.
The freedom to redefine $(\g^A,b^I)$ is due to the fact that we can change the dimensions of the elementary component fields,
as long as we keep the dimensions of the composite operators unchanged. 
This ambiguity is, therefore, not physical.
The physical $\beta$ function and the anomalous dimension matrices of the composite operators, defined in eq. 
\eqref{eq_anomalous_dimension}, are insensitive to this ambiguity.

Due to this ambiguity, even in theories where the holomorphic $\beta$-functions vanishes, there could be, in principle, parameterization of the equation in which the $b$ is linear in the couplings. This is still in agreement with our non-renormalization theorem of section \ref{sec_R_constraints}.  
Finally, let us comment that in the general formulation of the local RG equation  
there is an ambiguity which is similar in form but of different origin.  
As was demonstrated in \cite{Fortin:2012hc}, in certain cases there is a freedom to define a phase difference between the bare and renormalized sources.
This ambiguity in the definition of the bare Lagrangian leads to an ambiguity in the $\beta$ function.
However, this artifact does not appear in the holomorphic scheme we are using, in which the normalization of the wavefunction is kept as an independent field.

\subsection{Comparison with the general formulation}
\label{sec_components}

The original general expression for the \LRG~equation in the presence of scalar marginal operators, and vectors of dimension 3, is given by
\bea
\label{eq_Delta_W}
\Delta_{\sigma}^{W}&=&\int d^4x  
\le 2 \sigma g_{\mu\nu}\frac{\delta}{\delta g_{\mu\nu}(x)}
+\sigma \beta^I\frac{\delta}{\delta \lambda^I(x)}
+\le \sigma \rho^A_I\n_\mu\lambda^I - \d_\mu\sigma S^A\ri \frac{\delta }{\delta A_\mu^A(x)}
\ri
\eea
where $g_{\mu\nu}$ is the metric, $A^A_\mu$ are the background gauge fields, while $\rho$ and $S$ are some covariant functions of the sources $\lambda$ \cite{Osborn:1991gm}.

An important observation is that in this formalism there is no source for the kinetic term,
which corresponds to choosing canonical wave function renormalization.
In the superspace formulation such a term exists (related to the lowest component of $\Gamma$), 
and therefore the matching of the Ward identities in the two formulations is non-trivial. 
However, limiting ourselves to the operator equation,  i.e. \eqref{eq_operator_equation}, 
we can eliminate the contribution of the kinetic term, and compare the resulting equations.
For this purpose we focus on a specific point in parameter space, which is defined by the specific value $\tilde \lambda$ for the  coupling constants. 
We then use the ambiguity discussed in section \ref{sec_ambiguity} to subtract from $\Gamma^A$ the numerical value of its lowest component at that point, which we denote by $\wt \Gamma \equiv \Gamma( \tilde \lambda,\bar{\tilde \lambda})$  (it is not possible to subtract $\Gamma^A$ as a function, because it is not holomorphic):
\bea
\Delta^{\sW}_\s\left (b^I,\Gamma^A\right)\WW \Big |_ {\l=\tilde \l}&=& \Delta^\sW_\s\left (b^I+\wt \Gamma^A(T_A\l)^I,(\Gamma^A-\wt \Gamma^A)\right)\WW \Big |_ {\l=\tilde \l}+\Delta^G_{\Lambda =i \s \tilde{\Gamma}}\WW \Big |_ {\l=\tilde \l}~.\nl
\eea
Not surprisingly, the elimination of the wave function renormalization leads to the appearance to the physical $\beta$-function $\beta^I=b^I+\wt \Gamma^A(T_A\l)^I$. 

Another step necessary to make the comparison is to focus on the case where just the real part of the lowest component of the super-Weyl transformation parameter $\s$, which we denote by $\s_c$, is non zero.
To keep things simple, just the lowest component of the chiral coupling $\lambda$ is taken non-zero, and $e^V$ is written in the WZ gauge:
\bea
\l^I &=& \l_c^I+ i \t  \s^\mu \bt \d_\mu \l_c^I + \frac{\t^2 \bt^2}{4} \Box \l_c^I \nl
(e^V)_{\bi j} &=& \delta_{\bi j} +2 \theta \sigma^\mu \bar{\theta}  (A_\mu)_{\bi j} -
 \theta^2 \bar{\theta}^2 \le (A_\mu A^\mu)_{\bi j}+m^2_{\bi j} \ri
 \eea
 where the $D$-term of $V$ (the soft mass) is denoted by $m^2$.
Expanding the generator of the \sW~symmetry in components, and comparing with \eqref{eq_Delta_W} we find 
\beq
\beta^I = b^I +  \Gamma^A(T_A\l)^I \, , \qquad
\rho_I^A=-\d_I\Gamma^A \, , \qquad
\rho_\bI^A=\d_\bI\Gamma^A \, , \qquad
S^A=0 \, ,
\eeq
consistent with the results of \cite{Fortin:2012hc}. A non-vanishing $S^A$ function could, in principle, appear if $\Gamma$ had an anti-hermitean components. The vanishing of $S^A$ is, therefore, related to the $R$-symmetry constraints discussed in section \ref{sec_R_constraints}. 
This relation between $R$-symmetry and the vanishing of $S$ was discussed in \cite{Fortin:2012hc}\footnote{Notice, however, that this constraint is an artifact of the holomorphic scheme. It is not a direct consequence of supersymmetry. When working in components one can choose a scheme in which $S^A$ is non-zero.}.

 As explained in \cite {Osborn:1991gm}, in the presence of dimension two operators there could be contributions to the local RG equation which are non-vanishing in the limit where all the mass parameters are set to zero. In the SUSY case, one type of such contributions is related to holomorphic dimension 2 operators. As mentioned above, we chose to leave the discussion of such terms for future study. Another kind of dimension 2 operators are the soft SUSY-breaking mass term 
for the scalars in a chiral multiplet  (sourced by the $D$-term of $e^{V}$): 
\bea
\Delta_{\sigma}^{W}&\supset& 
-\int d^4x \sigma\le D_I^a\n^2 \lambda^I +  E_{IJ}^a\n_\mu \lambda^I \n^\mu \lambda ^J\ri\frac{\delta}{\delta (m^2)^a} \, ,
\eea
where in the case of the Wess-Zumino model
the index $a$ can be replaced by the product of a fundamental and anti-fundamental
indices $i \bj$:
\beq
D_I^{i \bj}=-\d_I\Gamma^A (T_A)^i_j \delta^{j \bj} \, , \qquad
E_{IJ}^{i \bj}=-\d_{IJ}\Gamma^A   (T_A)^i_j \delta^{j \bj} \, .
\eeq

 \subsection{Consistency conditions}
More constrains on $b$ and $\Gamma$ can be derived from the commutation relation of the generator of \sW~symmetry with the generators of flavor symmetries
\bea
[\Delta_{\s}^\sW,\Delta_{\Lambda}^G]&=&0~.
\eea
This condition implies that $b^I$ and $\Gamma^A$ are covariant functions of the sources, namely
\bea
\Delta_\Lambda^G ~b^I&=&i\Lambda^A (T_A b)^I\nl
\Delta_\Lambda^G ~\Gamma^A&=&i\left (\left [\Lambda,\Gamma\right]\right )^A~.
\eea

Another constraint can be derived from the commutation relation of the super-Weyl symmetry generator with itself
\bea
[\Delta_{\s_1}^\sW,\Delta_{\s_2}^\sW]&=&0~.
\eea
Imposing the vanishing of terms proportional to $\sigma_1\bar\sigma_2 - \sigma_2\bar\sigma_1$ in this commutator, we find the equation
\bea
\label{eq_BdotP}
\bab^\bI\d_\bI (e^V \Gamma)&=& \beta^I\d_I (\bar \Gamma e^V)~.
\eea
This constraint is used extensively in the computations described below. When written in components, this constraint agrees with the \LRG~result
\bea
\beta^I\rho_I^A+\bar\beta^\bI\rho_\bI^A&=&0~.
\eea

 \subsection{Super-Weyl variation of functions}

In this section we introduce some notations and terminology which facilitate the computation of Weyl variations of functions, 
and the analysis of the Weyl consistency conditions.

\subsubsection{SUSY covariant derivative}
To ensure the covariance of the super-Weyl anomaly under the global symmetries 
it is necessary to define a covariant supersymmetric derivative which involves the background gauge fields
\bea
\label{eq_covariant_derivative}
\N^\a \l^I =\mathcal D^\a\l^I + (A^\a)_J^I\l^J~,
\qquad
\bN^\da \bl^\bI = \bar{\mathcal{D}}^\da \bl^\bI + (B^\da)_\bJ^\bI \bl^\bJ~,
\eea
with $\mathcal D^\a$ the standard chiral covariant derivative in the curved background, and we introduce the notation
\bea
(A^\a)_J^I = (e^{-V})^{I\bar K} (\mathcal D^\a e^V)_{\bar K J}~,
\qquad
B^{\da \bI}_\bJ =   (\bDm^\da e^V)_{\bJ K}  (e^{-V} )^{K \bI }~.
\eea
The derivative $\N^\a$ ($\bN^\da $) has no connection part including the $\bI$ ($I$) indices;
for example
\beq
(\N^\a \xi)^{I J \bI}_{K \bJ}=\Dm^\a \xi^{I J \bI}_{K \bJ} + A^{\a I}_S  \xi^{S J  \bI}_{K \bJ} 
+  A^{\a J  }_S  \xi^{I S \bI}_{K \bJ}
-A^{\a S}_K \xi^{I J \bI}_{S  \bJ}  \, .
\eeq

The chiral field strength associated with the background gauge fields is given by
\bea
(W^\a)_J^I  = -\frac{1}{4} (\bDm^2 -4 R) (A^\a)_J^I  \, , \qquad
(\bar{W}^\da)_\bJ^\bI  = -\frac{1}{4} (\Dm^2 -4 \bar{R}) (B^\da)_\bJ^\bI \, .
\eea
Sometimes it is convenient to express $(W^\a)_J^I$
in terms of adjoint indices
\bea
(W^\a)_J^I  =(W^\a)^A (T_A)_J^I \, .
\eea
When writing in components we find the following useful result
\bea
\int d^2\theta \varphi^3 \text{Tr}[W^\a W_\a] \Big| _{\theta^2}+ c.c. \supset -4 \text{Tr}[F^{\mu\nu}F_{\mu\nu}] ~,
\eea
where $F^{\mu\nu}$ is the ``canonical'' field strenght, see footnote \ref{footnot}.

A useful property satisfied by the covariant derivative is $\N^\a (e^V)_{\bJ K}=\bN^\da (e^V)^{K \bJ} = 0$.
One of the consequences of this identity is that for any arbitrary covariant function of the sources  $Y=Y(\l ,\bl, e^V)$ we have
\bea
\N^\a Y =  \d_I Y\N^\a \l^I \, , \qquad
\bN^\da Y=  \d_{\bar I} Y\bN^\da \bl^ {\bar I} \, .
\eea

\subsubsection{Lie derivative}
\label{sec_Lie_derivative}
In order to express the super-Weyl variation of covariant functions in parameter space in a compact form,
it is convenient to define the following operators which we refer to as the Lie derivatives along the RG flow:
\bea
\li(Y_I)=\beta^J\d_J Y_I + \gamma_I^J Y_J
&\qquad&
\li(Y_\bI)=\beta^J\d_J Y_\bI + \bar{\wt \gamma}_\bI^\bJ Y_\bJ\nl
\bli(Y_I)=\bab^{\bar J}\d_{\bar J} Y_I + \wt \gamma_{I}^{J} Y_J
&\qquad&\bli(Y_\bI)=\bab^{\bar J}\d_{\bar J} Y_\bI + \bar{  \gamma}_{\bI}^{\bJ} Y_\bJ
\eea
where $Y_I$ is some arbitrary function of the sources, and the anomalous super-Weyl weight matrices $\gamma$ and $\tilde \gamma$ are defined in eq. \eqref{eq_anomalous_dimension}. The generalization to tensors with more indices is given in the appendix.
The Lie derivatives appear in the variation of functional involving the covariant derivative of couplings:
\bea
\Delta^{SW}_\s (Y_I \N^\a \l^I)
&=&\big (\frac 1 2 \s-\bs\big ) Y_I\N^\a \l^I + \Dm^\a \s ~Y_I \b^I + \s \li (Y_I)  \N^\a \l^I
+\bs   \bli (Y_I)  \N^\a \l^I \nl
\Delta^{SW}_\s(Y_\bI \bN^\da \bl^\bI)&=& 
\big (\frac 1 2 \bs-\s\big ) Y_\bI \bN^\da \bl^\bI + 
 \bDm^\da \bs ~Y_\bI \bab^\bI
  + \s \li (Y_\bI)  \bN^\da \bl^\bI
+\bs   \bli (Y_\bI)  \bN^\da \bl^\bI~~
\eea
(The first term in each line corresponds to the classical super-Weyl variations of the covariant derivative).

\subsubsection{Functions transforming covariantly under super-Weyl symmetry}
The transformation of higher order covariant derivatives has a more complicated form, e.g.:
\bea
\Delta_\s^{SW}(Y_I\N^2 \l^I)&=&
( \s-2\bs ) Y_I \N^2 \l^I +  \Dm^2 \s ~Y_K \b^K + 2 \Dm^\a \s   ~Y_K(\delta_I^K +\d_I\beta^K) \N_\a \l^I \nl
&&+ \s \mathcal{L}(Y_I) \N^2 \l^I
+\s Y_K\p_{IJ } \b^K \N^\a \l^I \N_\a \l^J \nl
&&+\bs\bar{\mathcal{L}}(Y_I)  \N^2 \l^I + 
\bs   Y_K \p_{IJ} ((e^{-V} \bg e^{V})^K_L \l^L )     \N^\a \l^I \N_\a \l^J 
\eea
As it was shown in the non-SUSY case \cite{Baume:2014rla}, it is sometimes convenient to use the following functions of the sources
\bea
\label{eq_Lambda_Pi}
\Lambda^I &=& (U^{-1})^I_K (\N^2 \l^K +4 \beta^K \bar{R}) \nl
\Pi^{IJ}&=&\N^\a \l^I \N_\a \l^J -\frac{1}{2} (\b^I \Lambda^J+\beta^J \Lambda^I) 
\eea
with $U_I^J=\delta_I^J +\d_I\beta^J$, in which some of the the derivatives of $\sigma$ appearing in the \sW~variation cancel, leading to the following \sW~transformations:
\bea
\label{eq_Lambda_Pi_transformation}
\Delta^{SW}_\sigma (Y_I\Lambda^I)
&=& (\s-2\bs)Y_I\Lambda^I +  2 Y_I  (\Dm^\a \s) \N_\a \l^I + \s \li(Y_I) \Lambda^I
+\bs \bli(Y_I)  \Lambda^I\nl
&&+\sigma Y_K \gamma^K_{IJ}\Pi^{IJ}+\bs Y_K  {\gamma}^K_{IJ} \Pi^{IJ} \nl
\Delta^{SW}_\sigma (Y_{IJ} \Pi^{IJ})
&=&  (\s-2\bs)Y_{IJ}\Pi^{IJ} +\s (\li(Y_{IJ})- Y_{KL}\gamma^{KL}_{IJ} ) \Pi^{IJ}
+\bs \le \bar{\mathcal{L}}(Y_{IJ}) -Y_{KL}   {\gamma}^{KL}_{IJ} \ri  \Pi^{IJ} \nl
\eea
The explicit expressions for the ${\gamma}^K_{IJ}$ and $  {\gamma}^{KL}_{IJ} $ 
tensors are given in eq. \eqref{eq_anomalous_dimension_tensors}.

 \section{Consistency conditions for the generalized super-Weyl anomaly}
\label{sec_anomaly}

 \subsection{The main idea}
   \label{sec_anomaly_idea}

As explained above, the superspace local RG (\sLRG) equation is an anomalous  Ward identity for the super-Weyl (\sW) symmetry off-criticality
\bea
\Delta_\sigma^\sW\WW&=&
\AA_\s ^\sW~.
\eea
Now that we have defined and discussed the generator of the super-Weyl symmetry $\Delta_\sigma^\sW$, it is time to move on to the right hand side of the equation and construct the anomaly. The anomaly encodes contact terms in the \sW~Ward identities for correlation functions, or equivalently, as explained in \cite{Baume:2014rla}, it contains information about the 
log divergences in the effective action $\WW$ and the corresponding counterterms (see eq. \eqref{eq_Lagrangian_W}).
The anomaly is parameterized by a set of 
functions which we refer to as the anomaly coefficients. These coefficients satisfy a non-trivial set of differential equations, due to the Wess-Zumino condition
\bea
\label{eq_Weyl_WZ_condition}
\int d^8z E^{-1}\delta^{WZ}_{\s_1,\s_2}&\equiv&\Delta^\sW_{\s_2} 
\AA_{\s_1} ^\sW-\Delta_{\s_1}^\sW 
\AA_{\s_2} ^\sW=0 \, .
\eea
The derivation of this set of equation is the main goal of this paper.

The procedure we follow consists of the following steps:
\begin{enumerate}
\item Write $\AA_\s^\sW$ as the most general scalar constructed from the sources, invariant under the global symmetries, with the correct classical \sW~weight (in section \ref{sec_chiral_anomalies} we show that in the presence of chiral anomalies it is also necessary to add some special terms which are not covariant under the global symmetries). In the supersymmetric case it is a non-trivial step to verify that one has the complete basis of terms in the anomaly.

\item
The generating functional $\WW$ is defined up to local terms, therefore the anomaly function is not-unique.
It is a crucial step to check how the different anomaly coefficients are modified under such changes of scheme, 
and identify the components of the anomaly which cannot be eliminated.

\item 
Compute the \sW~ variation of the anomaly, and write eq. \eqref{eq_Weyl_WZ_condition} in the following form
\bea
\label{eq_delta_WZ}
\delta^{WZ}_{\s_1,\s_2}
&=&
\s_{[1}\mathcal D^\a \s_{2]} \sum _af^1_a \mathcal S^a_{\a}+
\s_{[1}\bs_{2]} \sum_a f^2_a \mathcal S^{a}
+ \N^\a \s_{[1}\bN^\da \bs_{2]} \sum_a f^3_a \mathcal S^a_{\a\da}
+\N^2 \s_{[1}\bN^2 \bs_{2]}f ^4\nl
\eea
where $\mathcal S^a$ are an independent set of functions of space-time derivatives of the sources, (e.g. $R \N_\a\l^i $)
and $f_a$ are functions of the anomaly coefficients and their derivatives with respect to the sources. 
The equations for the vanishing of each of the functions $f_a$ are the Wess-Zumino consistency conditions.

\item
In the supersymmetric case there is an additional subtlety, not appearing in the non-supersymmetric case, as a certain combination of terms in $\delta^{WZ}_{\s_1,\s_2}$ can be written as a total derivative, and can therefore be removed from (or added to) the consistency conditions.
This ambiguity in the consistency conditions is parameterized using a chiral 
superfield $\Upsilon_A$, and an unconstrained superfield $\Omega_I$
\bea
\delta^{WZ}_{\s_1,\s_2}\sim \delta^{WZ}_{\s_1,\s_2} + F_1(\Upsilon_A)+ F_2(\Omega_I)
\eea
where the following functions can be written as total derivatives
\bea
\label{eq_total_derivative_cc}
F_1(\Upsilon_A)
&=&
\s_{[1}\mathcal D^\a \s_{2]} \Upsilon_A W^A_\a \, ,
\nl
&=&
-\frac{1}{4}\bDm^2(\s_{[1}\mathcal{D}^\a \s_{2]} \Upsilon_A A^A_\a)
\nl
F_2(\Omega_I)
&=&
\s_{[1}\mathcal D^\a \s_{2]}\Big(
 \p_{\bJ \bK} \Omega_I   \bN_\da \bl^\bK  \bN^\da \bl^\bJ \N_\a \l^I 
 +  \p_\bJ \Omega_I \bN^2 \bl^\bJ \N_\a \l^I \nl
 &&~~~~~~~~~~~~~~~~~~~~~~~~~~~~
+  4 i \p_\bJ \Omega_I \bN^\da \bl^\bJ  \N_{ \a \da}  \l^I 
-4 (W_\a \l)^I \Omega_I
\Big)\nl
&=&
\bDm^\da \le \s_{[1}\mathcal D^\a \s_{2]}\bDm_\da ( \Omega_I \N_\a \l^I)  \ri -\bDm_\da \le\bDm^\da \commuB ( \Omega_I \N_\a \l^I) \ri  ~.
\eea

\end{enumerate}
 \subsection{The super-Weyl anomaly}
   \label{sec_basis}
The anomaly can be found by writing the most general scalar function of the sources, allowed by symmetry and dimensional analysis\footnote{By dimensional analysis we mean that the anomaly terms must be classically invariant under a global rescaling.}, and then imposing the consistency condition.
In a supergravity background without the background fields $\lambda$ and $V$ one finds \cite{Bonora:1984pn}
\bea
\AA_\s^\sW\big|_{\l=0,V=0} &=& \int d^6 z \varphi^3  \sigma ~ \kappa W^{\alpha \beta \gamma}W_{\alpha \beta \gamma}
+ {\rm c.c.} 
\nl
&&- \int d^8 z E^{-1}  \left (\sigma  \left ( 2 a (G^2 +2R\bar R) + b R\bar R\right)  + D^2 \s d R\right)+{\rm c. c.}
\eea
where $W_{\alpha \beta \gamma}$, $G_{\a\da}$ and $R$ are the supergravity multiplets (we use the notations of  \cite{Buchbinder:1995uq};
see \cite{Bonora:2013rta} for the expression in Wess and Bagger conventions).
 Also, $G^2 = G^a G_a =-\frac 1 2 G^{\a\da}G_{\a\da}$), and the
 coefficients $\kappa$, $a$, $b$ and $d$ are model 
 dependent numbers\footnote{
Our normalization is such that $T^\mu_\mu=a E_4 - c W^2$, e.g.
 for a chiral multiplet  $a=\frac{1}{12 \times 64 \pi^2 } $, $c=\frac{1}{ 6 \times 64 \pi^2 }$.}
  (the non-standard choice of notations for $\kappa$ instead of $2(c-a)$ will be justified below).
 In fact, one can use the Wess-Zumino consistency condition \eqref{eq_Weyl_WZ_condition} to show that $b$ must vanish at the fixed point. 
Moreover, as we show below, by adding a local term to $\WW$ one can set $d$ to zero. 
 
In the presence of the background sources we can write many more terms. 
We find that a relatively convenient basis is the following:
\bea
\AA_\s^\sW &=&\AA_\s^\sW\big|_{\l=0,V=0}  +\AA_\s^{GW}
+ \int d^6 z \varphi^3  \sigma  \kappa_{A B} W^{A \a} W^B_\a  +
 {\rm c.c.} 
\nl
&&- \int d^8 z E^{-1} \sigma \Big( \chi_{I\bJ}G^{\a\da}\N_\a\l^I\bN_\da \bl^\bJ\nl
&&
~~~~
+\xi^1_{[IJ] \bK}   \N_{\a \da} \l^I \N^\a \l^J \bN^\da \bl^\bK
+\xi^2_{[\bI \bJ] K}  \N_{\a \da} \bl^\bI \bN^\da \bl^\bJ \N^\a \l^K
\nl
&&
~~~~
+ \zeta^1_{A I} W^{A\a} \N_\a \l^I + \zeta^2_{A \bI} \bar{W}^A_\da  \bN^\da \bl^\bI
\nl
&&
~~~~
+\epsilon^{1}_{I \bJ} \Lambda^I \bar \Lambda^\bJ
+\epsilon^{2}_{IJ \bK}   \Pi^{IJ} \bar \Lambda ^\bK
+\epsilon^{3}_{\bI \bJ K} \bN_\da \bl^\bI \bN^\da \bl^\bJ  \Lambda^K
 +\epsilon^4_{IJ \bK \bar L} \Pi^{IJ}\bar \Pi^{\bK\bar L}\nl
&&
~~~~
  +\eta^{1}_I R  \N^2 \l^I  
  +\eta^{2}_\bI   \bar{R}  \bN^2 \bl^\bI
+\eta^{3}_{IJ} R \N^\a \l^I \N_\a \l^J  
  +\eta^{4}_{\bI \bJ}  \bar{R} \bN_\da \bl^\bI \bN^\da \bl^\bJ  \Big)
+{\rm c. c.}\nl
&&-  \int d^8 z E^{-1} \mathcal D^\a\sigma \Big(
 w^1_\bI G_{\a\da}\bN^\da \bl^\bI+ w^2_I R\N_\a \l^I
+u_{I\bJ}^1 \N_\a \l^I \Lambda^\bJ 
+u_{I \bJ\bK}^2 \N_\a \l^I \N_\da \bl^\bJ \bN^\da\bl^\bK \Big)
+{\rm c. c.}
\nl
&&
- \int d^8 z E^{-1} \mathcal D^2\sigma \Big(
 v^1_\bI \bar \Lambda^\bI
+
 v^2_{\bI\bJ} \bar \Pi^{\bI\bJ}\Big)
+{\rm c. c.}
\label{big_anomaly}
\eea
$\Lambda$ and $\Pi$ are the functions defined in \eqref{eq_Lambda_Pi}, and $\N_{\a\da}=\frac{i}{2}\{\N_\a,\bN_{\da}\}$. 
The term $\AA_\s^{GW}$
is a non-covariant anomaly term required by the consistency with chiral anomalies
 and will be discussed in the next section.
Let us make some comments regarding this basis:
\begin{itemize}
\item In the presence of the background sources, the anomaly coefficients are generalized to be covariant functions of the sources (the covariance is implied by the fact that Weyl symmetry commutes with the global symmetries).

\item 
There are terms which seem to be missing from this basis (such as $\N_{\a\da}\l\N^{\a\da}\l$), but we have verified that all these terms can be integrated by parts and absorbed in the terms appearing in this formula. 
The approach we were following was to start with a basis where the derivatives are acting only on the sources (and not on $\sigma$), and then integrating by parts when necessary.

\item The matching between the superfield anomaly coefficients and the anomaly coefficients in the \LRG~anomalies is sometimes non-trivial. Each of the superWeyl anomalies, written in components, contains several anomaly terms with space-time derivatives of $\sigma$ and the anomaly coefficients themselves.

\item We chose to write some of the anomalies in the basis of the functions $\Lambda$ and $\Pi$, in order to simplify the consistency conditions. It is useful, however, to identify the linear combination of anomaly coefficients which correspond to the $\N^2\l^I\bN^2\bl^\bJ$ anomaly:
\bea
\label{eq_g_IJ}
g_{I\bJ}&\equiv&
( {U}^{-1})^K_{I}(\bar {U}^{-1})^\bbL_{\bJ}(\epsilon^{1}_{K\bbL}-\epsilon^{2}_{KM\bbL}\beta^M+
\epsilon^{4}_{KM\bbL\bar N}\b^M\bab^{\bar N})
\label{def_metrica_zamolo}
\eea
At the fixed point, the hermitean part of this matrix
 (specializing to components corresponding to exactly marginal primary operators), 
 is proportional to the Zamolodchikov metric \cite{Zamolodchikov:1986gt}
\bea
\mathcal G_{I\bJ} &\equiv&  \langle \OO_I(0)\bar \OO_\bJ (z)\rangle ( z^2)^3 
\label{zamolo}
\eea
The relation between the metric and the super-Weyl anomaly coefficients is clarified in appendix \ref{app_metric}.

\item  We have not included in the anomaly terms such as
\beq
\int d^8 z E^{-1} \s (\alpha R^2 + \alpha_I \bar{R} \N^2 \l^I + \alpha_{IJ} \N^2 \l^I \N^2 \l^J )~,
\eeq
whose global Weyl variation is proportional to $(\s -\bs)$. 
We checked by direct calculation that the
consistency condition for these terms decouple from the ones
for the other terms (which include the interesting ones which are related 
to the central charge $a$). In this paper we will not discuss these
terms and we will leave this as a topic for further investigation. 

\item There was a previous attempt \cite{Grosse:2007au}
 to write the Wess-Zumino
consistency conditions for the anomaly in eq.~(\ref{big_anomaly}),
which missed many of the crucial ingredients, such as the background gauge fields.

\end{itemize}

 \subsection{The super-Weyl anomaly in the presence of chiral $U(1)$ anomalies}
   \label{sec_chiral_anomalies}
In addition to the WZ consistency condition \eqref{eq_Weyl_WZ_condition}, the \sW~anomaly must satisfy another constraint
\bea
\label{eq_Weyl_flavor_WZ_condition}
\Delta_{\Lambda}^G 
\AA_{\s} ^\sW
-
\Delta^\sW_{\s} 
\AA_{\Lambda} ^G
&=&0
\eea
which is the WZ condition associated with the fact that the super-Weyl symmetry commutes with the global symmetries.  This constraint was first discussed in the non-SUSY case in \cite{Keren-Zur:2014sva}.
If the global symmetries are anomaly free ($\AA_\Lambda^G=0$), then this constraint is satisfied if $\AA^\sW_\s$ is invariant under the internal symmetries.
If the theory does contain chiral anomalies, there is another consistent possibility in which the chiral anomaly $\AA_\Lambda^G$ is \sW~invariant (which is the case of the chiral-gravitational anomaly).

In general, however, the chiral anomaly is not \sW~invariant, due to the \sW~transformation of the background gauge fields defined by \eqref{eq_Delta_W}. This implies that the condition \eqref{eq_Weyl_flavor_WZ_condition} can be satisfied only if we add terms to the Weyl anomaly which are not singlets of the anomalous flavor symmetry. An important observation is that  $\AA_\Lambda^G$ is the \emph{consistent} chiral anomaly, and not the \emph{covariant} one. 
While in the case of anomalies involving only abelian currents, the consistent and covariant anomalies are identical (up to an overall numerical factor), in the non-abelian case the consistent anomaly does not have a closed local form, and is given only in terms of an integral over an auxiliary parameter (see e.g. \cite{Guadagnini:1985ar,McArthur:1985xd,Marinkovic:1990ny,Ohshima:1999jg,Gates:2000dq,Gates:2000gu}).
In order to simplify the analysis here, we will focus on theories with only $U(1)$ currents involved in the chiral anomalies.
We leave the computation of the Weyl anomaly in the presence of anomalous global non-abelian currents for future work.

The consistent abelian anomaly is given by
\bea
\label{eq_consistent_anomaly}
\AA_\Lambda^G &=& 
\int d^6z\varphi^3~ i\Lambda^A \le 
k_A W^{\a\b\gamma}W_{\a\b\gamma}+k_{ABC}W^{\a B}W_\a^C  \ri  + c.c~,
\eea
where $W^{\a\b\gamma}$ is the Weyl tensor, $W^{\a A}$ is the field strength associated with the gauge field $V^A$, and the coefficients are given by
\bea
k_{A}\equiv -\frac 1 {192\pi^2} {\rm Tr} \left [ T_A\right]
\qquad \qquad
k_{ABC}\equiv -\frac 1 {192\pi^2} \frac{1}{2} \rm {Tr} \left [ T_A \{T_B, T_C\} \right]~.
\label{eq_k_ABC}
\eea
$k_{ABC}$ is $\frac 1 3$ times the coefficient of the covariant anomaly.
The traces in~\eqref{eq_k_ABC} are taken on generators
acting on the elementary fermionic field content of the theory.
In the absence of holomorphic singlet functions, as discussed in section \ref{sec_flavor_symmetry}, the expressions for $k_A$ and $k_{ABC}$ given here are the only tensors with the right index structure that one can write. This can be viewed as a quick derivation of the Adler-Bardeen theorem \cite{Adler:1969er} for supersymmetric theories which satisfy this condition.

The solution to the Wess-Zumino consistency condition \eqref{eq_Weyl_flavor_WZ_condition} is found using the following non-trivial identity:
\bea
\Delta_\s^\sW \AA_\Lambda^G&=&-\Delta_\Lambda^G \Big (\int d^8z E^{-1} (\s +\bs)\Gamma^A X_A\Big)~,
\eea
where $X_A$ is a non-covariant function of the background gauge fields:
\bea
X_A&=& 2k_{ABC}
\left (\N^\a (V^BW_{\a}^{ C}) +\bN_\da(V^B \bar W^{\da C})- \frac 1 2 V^B(\N^\a W_{\a}^{ C} + \bN_\da \bar W^{\da C})\right)~.
\eea
$X^A$ is the superfield analog of the function found in \cite{Bardeen:1984pm}, which gives the difference between the consistent and covariant abelian currents. 
This identity suggests that eq. \eqref{eq_Weyl_flavor_WZ_condition} is satisfied only if we add the following term to the Weyl anomaly
\bea
\AA_\sigma^{W} 
&\supset& \AA_\s^{GW}\equiv-\int d^8z E^{-1}(\s+\bs)\Gamma^A X_A~.
\eea
This is similar to the solution found in \cite{Keren-Zur:2014sva}, although we have to stress again, that we checked its validity in the \sLRG~equation only for anomalies involving abelian symmetries.

The anomaly term found here is not covariant, and can be rewritten as
\bea
\label{eq_new_Weyl_anomaly}
 \AA_\s^{GW}&=& 2k_{ABC}\int d^8zE^{-1}~  \sigma \Big(
  \Gamma^C  \N^\a V^A W_{\a}^{ B}
   -\d_\bi \Gamma^C V^A  \bar W_\da^ B  \bN^\da \bl^\bi)
\Big)+c.c.~~\eea
The first term has the same form as the consistent anomaly \eqref{eq_consistent_anomaly}, but since the function $\Gamma$ is not holomorphic, this anomaly is not covariant.

A simple interpretation of the new anomaly and the new functions can be found by considering the Ward identity
 in the presence of the background sources
\bea
[\TT]
&=&- b^I [\OO_I] -\frac 1 4 \Gamma^A \bN^2 [J_A]-\frac 1 4 \Gamma^A \bN^2 X_A +\ldots
\eea
We see that the new anomaly can be absorbed by replacing the consistent current $J_A$ with the covariant one $\tilde J_A$
\bea[\tilde J_A]\equiv [J_A]+X_A~.
\eea

\subsubsection{Chiral anomalies, ambiguity and Weyl anomaly coefficients}
The appearance of the chiral anomalies implies that the ambiguity discussed in section \ref{sec_ambiguity}, associated with a redefinition of the \sW~generator
\bea
\Delta_\s^\sW \to 
(\Delta_\s^\sW )'=\Delta_\sigma^\sW +\Delta_{\Lambda=ia\sigma}^G~,
\eea
is valid only if we also shift the anomaly
\bea
\AA_\s^\sW \to (\AA_\s^\sW )'=\AA_\sigma^\sW +\AA_{\Lambda=ia\sigma}^G~.
\eea
Under this transformation, the \sW~anomaly coefficient $\kappa$ and $\kappa_{AB}$ are shifted as\footnote{In order to find the shift in $\kappa$ one has 
 to first combine the $\kappa$ anomaly and the first term in eq. \eqref{eq_new_Weyl_anomaly} into one non-holomorphic anomaly coefficient $\kappa_{AB} + 2d_{ABC} \Gamma^C$, and take into account the transformation of $\Gamma$.}
\bea
 \kappa&\to& \kappa'=\kappa - k_Aa^A\nl
 \kappa_{AB}&\to& \kappa'_{AB}  = \kappa_{AB}- 3k_{ABC} a^C
\eea
We can therefore a define non-ambiguous functions 
\bea
\label{eq_tau}
\tau_{AB}&=&\kappa_{AB}+ 3k_{ABC} \Gamma^C\nl
2(c-a)&=&\kappa +  k_A \Gamma^A~.
\eea
 These functions appear in the \sW~Ward identity after the currents are eliminated using a flavor Ward identity
as in eq. \eqref{eq_operator_equation_beta} (now the covariant Ward identity has to be used):
 \bea
[\TT]
&=&- b^I [\OO_I] -\frac 1 4 \Gamma^A \bN^2 [\tilde J_A]+
 \kappa W^{\a\b\gamma}W_{\a\b\gamma}+\kappa_{AB}W^{\a A}W_\a^B\nl
&=&- \beta^I [\OO_I]  + 2(c-a) W^{\a\b\gamma}W_{\a\b\gamma}+\tau_{AB}W^{\a A}W_\a^B+\ldots
\eea 
The expressions for $\tau_{AB}$ and $c-a$ are generalization of results appearing in \cite{Anselmi:1997am},
\cite {Anselmi:1997ys} and \cite{Babington:2005vu}.

At the fixed point, 
the imaginary part of the $\theta^2$ component of this equation
 gives the anomalous Ward identity of the $R$-symmetry.
 Using results from \cite{Anselmi:1997am,Anselmi:1997ys}, we find 
\bea
\label{eq_tau_fp}
2(c-a) \big |_{f.p} &=& -\frac{1}{128\pi^2}Tr[T_{R,f}]\nl
\tau_{AB} \big |_{f.p} &=& -\frac{3}{128\pi^2}Tr[T_{R,f}T_AT_B]~.
\eea
where $T_{R,f}$ are the $R$-charges of the fermions
and the trace is on
the microscopic field content space.

The $R$-charge can be written in terms of 
$\Gamma$ at the fixed point
\bea
T_{R,f} &=&  \frac 2 3 \Gamma^AT_A|_{f.p} - \frac 1 3~.
\eea
Combining this with eq. \eqref{eq_tau} and \eqref{eq_tau_fp} we find that 
\bea
\kappa|_{f.p}&=& \frac{1}{384\pi^2}Tr[1]\nl
\kappa_{AB}|_{f.p}&=&\frac 1{128\pi^2}Tr[T_AT_B] \, ,
\label{eq_k_AB}
\eea
(assuming we use the parameterization in which $T_R$ coincides with the superconformal $R$ symmetry). 
In theories with no chiral singlets, these expressions are valid off-criticality as well, since there is no chiral function of the sources with the necessary index structure. This property will be important in the discussion appearing in section \ref{sec_a_tilde}.
Finally, we comment that this result agrees with an interpretation of $\kappa_{AB}$ as the holomorphic $\beta$ function for a weakly gauged global symmetry.

 \subsection{Scheme dependence of the anomaly}
   \label{sec_scheme}
An important aspect of the anomaly which is relevant to our discussion is its dependence on the choice of scheme. 
By choice of scheme we refer here to the freedom to add local terms to the generating functional $\WW$. The super-Weyl variation of these local terms modifies the anomaly coefficients. For example, 
the anomaly coefficients $v^1$, $v^2$ and $u^2$, and the real part of $d$ and $u^1$
can be eliminated by adding the following local term 
\bea
\delta \WW &=&
 -\int d^8 z E^{-1}\left(4 \bar R \left(d R +  v^1_\bI \bar \Lambda^\bI+ v^2_{\bI\bJ} \bar \Pi^{\bI\bJ}\right)
 -\frac 1 2 \Lambda^I \left (u_{I \bJ}^1 \Lambda^\bJ
+u_{I \bJ\bK}^2 \N_\da \bl^\bJ \bN^\da\bl^\bK\right)
\right)+{\rm c. c.}\nl
\eea
(Notice that other anomaly coefficients are modified in this process).
In the following we assume that we work in such a scheme.

One can consider the effect of adding other counter terms. An important example is 
\bea
\delta \WW &=&
- \int d^8 z E^{-1}  2A   (G^2 +2R\bar R)
\eea
with some arbitrary function $A$. Under this action, the anomaly coefficients are modified as
\bea
a\to a +\beta^I\d_I A
\qquad
w^1_\bI\to w^1_\bI-  \d_\bI A
\qquad
w^2_I\to w^2_I+  \d_I A
\eea
We will later see see, by inspecting the consistency conditions,
that $w^1_\bI = -\bar{w}^2_\bI \equiv w_{\bI}$.
Using this result,
 we can define a function, which we denote by $\tilde a$, which is invariant under this transformation
\bea
\tilde a&\equiv&a+\bar{w}_I  \beta^I~.
\label{aaatilde}
\eea
This function will play an important role in our discussion. A useful observation is that $\tilde a$ is sensitive to the choice of scheme related to the following counter term:
\bea
\delta W &=& - \int d^8z E^{-1}
C_{I \bJ}G^{\a\da} \N_\a\l^I \bN_\da \bl^\bJ~.
\eea
Indeed, under the addition of this term $(a,\bar{w}_I,\tilde a)$ transform as
\bea
a \to a \, , \qquad 
\bar{w}_I \to C_{I \bJ} \bab^\bJ \, , \qquad
\tilde a \to\tilde a +  C_{I\bJ}\beta^I\bab^\bJ \, .
\label{aaaambiguity}
\eea
We conclude that any contribution to $\tilde a$ which is of order $\beta\bab$ can be eliminated by a choice of scheme.

 \subsection{Consistency conditions}
 \label{sec_consistency_conditions}
We are now in the position to give the consistency conditions, 
which are derived from the constraint discussed in section \ref{sec_anomaly_idea}.
We will write the various constraints $f^{1,2,3,4}=0$, associated with the various terms 
 $(\mathcal{S}^a, \mathcal{S}^a_\a,\mathcal{S}^a_{\a \da})$  as defined in 
 eq.~\ref{eq_delta_WZ}.
All the equations are written in the scheme described in section
 \ref{sec_scheme} in which we eliminate as many terms as possible.
In certain cases we write the equation already
taking into account constraints obtained from previous equations. 

\subsubsection{Constraints setting anomaly coefficients to zero}
Some of the constraints enforce the vanishing of certain anomaly coefficients:
\begin{center}
    \begin{tabular}{ c | | l l r }
    \hline
$ \commuA \bN^2\N^2 \l^I$
&
$~~\eta_I^1~=$&$0$
&
\parbox{0.1\textwidth}{\beq ~\eeq}
\\    \hline
$ \commuB \N_\a \bN^2 \bl^\bI$
&
$~~\eta_\bI^2~=$&$0$
&
\parbox{0.1\textwidth}{\beq ~\eeq}
\\    \hline
$ \commuA \N_{\a\da}\l^I \N^{\a\da}\l^J$
&
$~~\eta_{IJ}^3~=$&$0$
&
\parbox{0.1\textwidth}{\beq ~\eeq}
\\    \hline
$ \commuB \N^{\da}\bl^{(\bI} \N_{\a\da}\bl^{\bJ)}$
&
$~~\eta_{\bI\bJ}^4~=$&$0$
&
\parbox{0.1\textwidth}{\beq ~\eeq}
\\    \hline
$  \commuB \Dm_\a R$
&
$~~b~~~~~=$&$0$
&
\parbox{0.1\textwidth}{\beq ~\eeq}
\\    \hline
$  ( \N^2 \s_{[1}\bN^2 \bs_{2]})$
&
$~~d-\bar d~=$&$0$
&
\parbox{0.1\textwidth}{\beq ~\eeq}
\\    \hline
    \end{tabular}
\end{center}
  Another useful constraint we use at this stage is
  \begin{center}
    \begin{tabular}{ c | | ll r }
    \hline
$   \commuC \N_{\a \da} \l^I$
&
$~~w^2_I~=$&$-\bar w^1_I$
&
\parbox{0.1\textwidth}{\beq ~\eeq}
\\    \hline
    \end{tabular}
\end{center}
Using this information and the choice of scheme described above, the anomaly now takes the form
\bea
\AA_\s^\sW &=& \int d^6 z \varphi^3  \sigma \left( \kappa (W_{\alpha \beta \gamma})^2  \right) +
 {\rm c.c.} 
\nl
&&- \int d^8 z E^{-1} \sigma \Big(2 a (G^2 +2R\bar R) +  \chi_{I\bJ}G^{\a\da}\N_\a\l^I\bN_\da \bl^\bJ\nl
&&
~~~~
+\xi^1_{[IJ] \bK}   \N_{\a \da} \l^I \N^\a \l^J \bN^\da \bl^\bK
+\xi^2_{[\bI \bJ] K}  \N_{\a \da} \bl^\bi \bN^\da \bl^\bj \N^\a \l^k
\nl
&&
~~~~
+ \zeta^1_{A I} W^{A\a} \N_\a \l^I +
 (\zeta^2_{A \bI} +2 k_{ABC}\d_\bI  \Gamma^B V^C)\bar{W}^A_\da  \bN^\da \bl^\bI
 - (\kappa_{A B}+2 k_{ABC}\Gamma^C)  W^{A \a}A^B_\a
\nl
&&
~~~~
+\epsilon^{1}_{I \bJ} \Lambda^I \bar \Lambda^\bJ
+\epsilon^{2}_{IJ \bK}   \Pi^{IJ} \bar \Lambda ^\bK
+\epsilon^{3}_{\bI \bJ K} \bN_\da \bl^\bI \bN^\da \bl^\bJ  \Lambda^K
 +\epsilon^4_{IJ \bK\bar L} \Pi^{IJ}\bar \Pi^{\bK \bar L}\Big)
+{\rm c. c.}\nl
&&-  \int d^8 z E^{-1} \mathcal D^\a\sigma \Big(
 w_\bI G_{\a\da}\bN^\da \bl^\bI+ \bar w_I R\N_\a \l^I
+u_{[I \bJ]} \N_\a \l^I \Lambda^\bJ
\Big)
+{\rm c. c.}
\eea
where we use the notations $w_\bI \equiv w^1_\bI=-\bar w^2_\bI$ and  
$u_{[I \bJ]} \equiv \half (u_{I\bJ}^1-\bar u_{\bJ I }^1)$.

\subsubsection{Constraints which can be used to algebraically eliminate anomaly coefficients}
The following constraints can be used to write some anomaly coefficients
as an algebraic function of the others:
 \begin{center}
    \begin{tabular}{ c | | ll r }
    \hline
$\commuB \N_\a \l^I \bar{\Lambda}^\bJ$
&
$~~8\epsilon^1_{I \bJ}~=$&$
\bar{U}^\bK_\bJ  \le -\chi_{I \bK} +4 \li (u_{[I\bbL]} (\bar{U}^{-1})^\bbL_\bK ) \ri
$
&
\parbox{0.1\textwidth}{\beq ~\label{positivismo}\eeq}
\\    \hline
\tabspace 
$  \commuA \N_{\a\da}\l^{(I} \N^\a \l^{J)}\bN^\da\bl^\bK
$
&
$~~8\epsilon_{(IJ) \bK}^2~=$&$
-  \p_{(J} \chi_{I) \bK} 
+\mathcal{I}^0_{(IJ) \bK} 
 +\mathcal{I}_{(I \bK A}^{1}\d_{J)}\Gamma^A$
&
\parbox{0.1\textwidth}{\beq ~\label{weeq1} \eeq}
\\    \hline
\tabspace 
$  
 \commuB \N_\a \l^K  \bN_\da \bl^{\bI} \bN^\da \bl^{\bJ} 
$
&
$~~8\epsilon_{(\bI \bJ) K}^3~=$&$
-  \p_{(\bJ} \chi_{ K \bI)} 
+\bar{\mathcal{I}}^0_{(\bI \bJ) K} 
+ \mathcal{I}_{K(\bI A}^{1}\d_{\bJ)}\Gamma^A$
&
\parbox{0.1\textwidth}{\beq ~\label{weeq2}  \eeq}
\\    \hline
    \tabspace
$
\commuC \N_\a \l^I  \bN_\da \bl^{\bJ}$
&
$~~
8u_{[I \bJ]}~ =$&$ 
\d_\bJ \bar w_I - \p_I w_\bJ+ i   {\xi}^{2}_{[\bK\bJ] I} \bab^\bK+ i   \bar{\xi}^{2}_{[KI] \bJ} \beta^K  
$
&
\parbox{0.1\textwidth}{\beq ~\eeq}
\\    \hline
    \end{tabular}
\end{center}
where:
\bea
\mathcal{I}^0_{(IJ) \bK} &\equiv& 
 4 u_{[\bK L] } {{\gamma}}^L_{IJ}  + \bar \zeta^2_{A (I} \p_{J) \bK}\Gamma^A\nl
\mathcal{I}_{I\bJ A}^{1}&\equiv&
2i\xi^1_{[KI] \bJ} (T_A\l)^K
+2i\xi^2_{[\bK \bJ] I}(e^{V}T_Ae^{-V}\bl)^\bK   \nl
&&-\d_\bJ \zeta^{1}_{ IA} 
-\d_I \zeta^{2~\bbL}_{\bJ  \bK}( e^VT_A e^{-V})_\bbL^\bK~.
\label{iato1}
\eea
and the symmetrization in eqs.~(\ref{weeq1}, \ref{weeq2}) is just on the $(I,J)$
and $(\bI,\bJ)$ indices.
The symbol $\gamma^K_{IJ}$ defined in \eqref{eq_anomalous_dimension_tensors}
and $\zeta^{2\bbL}_{\bJ  \bK}$ is defined such that 
$\zeta^2_{\bJ A}\bar W^A_\da = \zeta^{2\bbL}_{\bJ \bK}(\bar W_\da)_\bbL^\bK$.

We are now left with the following independent anomaly coefficients:
\bea
a,~\kappa,~\kappa_{AB},~\chi_{I\bJ},~w_\bI,~\xi_{IJ\bK}^1,~\xi_{\bI\bJ K}^2, ~\zeta_{AI}^1, ~\zeta_{A\bI}^2,~\epsilon_{IJ\bK\bbL}^4
\eea

\subsubsection{Reality constraints}
Some of the consistency condition take the form of reality constraints
on some combinations of the anomaly coefficients:
  \begin{center}
    \begin{tabular}{ c | | l r }
    \hline
$   \commuC   G_{\a \da}
 $
&
$~~ \tilde a = \bar {\tilde a }
$
&
\parbox{0.1\textwidth}{\beq ~ \label{aaareal} \eeq}
\\    \hline
$  \commuA G^{\a \da} G_{\a \da} 
 $
&
$~~ \bli(a)=\li(\bar a)
$
&
\parbox{0.1\textwidth}{\beq ~\eeq}
\\    \hline
$  \commuA G_{\a \da}  \N^\a \l^i \bN^\da \bl^{\bj}
 $
&
$~~ \bli(\chi_{I\bJ})=\li(\bar \chi_{\bJ I})
$
&
\parbox{0.1\textwidth}{\beq ~\eeq}
\\    \hline
$    \commuA   \Lambda^I\bar \Lambda^\bJ
 $
&
$~~ \bli(g_{I\bJ})=\li(\bar g_{\bJ I})
$
&
\parbox{0.1\textwidth}{\beq ~\eeq}
\\    \hline
$ \commuA  \N^\a W^A_\a 
 $
&
$~~ (e^V)_{\bJ K} (\bar{w}_I\l^K+ \b^L\bar{\zeta}^{2~K}_{LI}) 
~=h.c.
$
&
\parbox{0.1\textwidth}{\beq \label{eq_beta_eta} ~\eeq}
\\    \hline
$   \commuA  \N^{\a \da} \l^I \N_{\a \da} \bl^\bJ $
&
$~~ 
  \chi_{I \bJ}- \p_I w_\bJ
-   2i \xi^{2}_{[\bJ\bK] I}  \bab^\bK  
  +  \zeta^2_{A \bJ} \p_I \bar \g^A
~= h.c.$
&
\parbox{0.1\textwidth}{\beq ~\label{eq_useful} \eeq}
\\    \hline
\tabspace$    \commuA \N^\a \l^I \N_\a \l^J \bN_\da \bl^\bK \bN^\da \bl ^\bbL
$
&
$
     \bli( \epsilon^4_{IJ\bK\bbL} )   -  \epsilon^4_{MN \bK \bbL} {\gamma}^{MN}_{IJ}  
-   \epsilon^4_{IJ \bar M \bar N}\bar {\gamma}^{\bar M\bar N}_{\bK \bbL}
+ \epsilon^{2}_{IJ \bar M } \bar{\gamma}^{\bar M}_{\bK \bbL}
$&\\&$
~~~~~~~~~~ +  \epsilon^{3}_{\bI \bJ M } {\gamma}^M_{KL} 
-\frac{1}{4} \mathcal{I}^1_{I \bK  A}\d_{  J \bbL} \Gamma^A ~~~=h.c.
$
&
\parbox{0.1\textwidth}{\beq ~\eeq}
\\    \hline
    \end{tabular}
\end{center}
where the quantities $\tilde{a}$ and $g_{I \bJ}$ 
are defined in eq.~\eqref{aaatilde} and eq.~\eqref{def_metrica_zamolo}.

\subsubsection{Other constraints}
 The following consistency conditions will be further discussed in section \ref{sec_implications};
 they give constraints on the $a$-function and on the conformal manifold:
  \begin{center}
    \begin{tabular}{ c | l r }
    \hline
$ \commuB \bN^\da \l^\bI G_{\a \da}$
&
$~~
\p_\bI a ~= \beta^J \chi_{J \bI}   - \li(w_\bI)$
&
\parbox{0.1\textwidth}{\beq \label{eq_gradient_flow}~\eeq}
\\    \hline
\tabspace$ \commuB W^A_\a $
&
$ ~~ -2 \kappa_{AB} \g^B -3k_{ABC}\Gamma^B\Gamma^C +  \zeta^1_{A I}  \beta^I 
+w_\bI (e^V T_A e^{-V}\bl)^\bI$
\\
&
~~~~~~=
$\Upsilon_A + \Omega _I (T_A\l)^I$
&
\parbox{0.1\textwidth}{\beq \label{amaxgeneralized} ~\eeq}
\\    \hline
$ \commuB  \N_{\a \da} \l^I \bN^\da \bl^\bJ  $
&
$~~  \chi_{I \bJ}- \p_I w_\bJ 
-   2i \xi^{1}_{[IK] \bJ}  \b^K  
  +  \zeta^1_{A I} \p_\bJ \g^A
~=
  - \p_\bJ \Omega_I$
&
\parbox{0.1\textwidth}{\beq ~ \label{chisolution} \eeq}
\\    \hline  
$    \commuB \N_{\a \da} \bl^\bI \bN^\da \bl^\bJ 
$
&
$ ~~ \p_{\bI} w_{\bJ} -\p_{\bJ} w_{\bI}
~=2 i \xi^{2}_{[\bI \bJ] K}  \b^K 
 + \le  \zeta^2_{A \bJ}  \d_\bI\bar \Gamma^A-\zeta^2_{A \bI} \d_\bJ \bar \Gamma^A  \ri 
$
&
\parbox{0.1\textwidth}{\beq ~\label{eq_rotore} \eeq}
\\    \hline  
    \end{tabular}
\end{center}
Combining \eqref{eq_gradient_flow} with other constraints 
eqs.~(\ref{aaareal},\ref{eq_beta_eta},\ref{eq_useful},\ref{eq_rotore}), one finds a similar equation
\bea
 \p_I a ~= \bab^\bJ  \chi_{I \bJ}   - \li(\bar w_I) \, .
 \label{eq_gradient_flow2}
\eea
The same equation is found in a completely different way
 from the $\commuA G_{\a \da} \N^{\a \da} \l^I$ consistency condition;
 this is a non-trivial cross-check of our calculations.
An interesting property of eq. \eqref{amaxgeneralized} is that is seems to be sensitive to the ambiguity discussed
in eqs.~(\ref{ambi1},\ref{ambi2}).
  However, the invariance of the equation can be restored by assigning $\Upsilon_A$ appropriate transformation properties:
  \beq
  \Upsilon'_A = \Upsilon_A -2 \kappa_{AB} a^B - 3 \kappa_{ABC} a^B a^C \, .
  \label{ambi3}
  \eeq

Finally, we list the set of remaining constraints,
which we will not use in the discussion of the physical implication:
  \begin{center}
    \begin{tabular}{ c | | lr }
  \hline
$     \commuA \N_{\a\da}\bl^{[\bI} \bN^\da \bl^{\bJ]}\N^\a\l^K $
&
$~~ 
  \p_{[J}  \chi_{I ]\bK} 
   =   i\li(\bar{\xi}^{2}_{[IJ] \bK})+
 \mathcal{I}^2_{IJ\bK\bbL}\bab^\bbL
+\mathcal{I}_{[J\bK A}^{1}\d_{I]}\Gamma^A
 - \bar{\zeta}^2_{A [I} \p_{J]\bK} \g^A
$
&
\parbox{0.1\textwidth}{
\beq\label{eq_almost_Kahler} 
 ~\eeq}
\\    \hline
$ \commuA W^{\a A} \N_\a \l^I $
&
$~~ \chi_{I \bJ }(e^V T_A e^{-V}\bl)^\bJ~=\li(\bar \zeta^2_{AI}) + \mathcal I^1_{I\bJ A}\bab^\bj -2 \tau_{A B} \d_I\Gamma^B
+\mathcal{I}^3_{A I}
$
&
\parbox{0.1\textwidth}{
\beq\label{eq_vector_beta_function}
 ~\eeq}
\\    \hline
  $    \commuA \N^\a \l^I \N_\a \bN^2\bl^\bJ
$
&
$~~8(\epsilon^1_{I \bJ}-\beta^K\epsilon^2_{IK\bJ} + \bab^\bK \epsilon^3_{\bJ\bK I })$\nl
&$~~~~~~~~~~~=
{U}^K_I  \le -\chi_{K\bJ} +4 \li (u_{[L\bJ]} ({U}^{-1})^L_K ) \ri$
&
\parbox{0.1\textwidth}{\beq ~\eeq}
\\    \hline
$   \commuA \N^\a \l^I \N_\a \l^J \bar{\Lambda}^\bK
$
&
$ 0~=  ~
\gamma^L_{IJ}  (\epsilon^1_{L \bK}  - \bar{\epsilon}^1_{\bK L } )
+\bli(\epsilon^2_{I J \bK}) - \epsilon^2_{L M \bK} \gamma^{L M}_{I J} 
$
&\\
& $~~~~
- \li(\bar{\epsilon}^3_{IJ \bK}) +\bab^{\bar{L}}  \bar{\epsilon}^3_{\bK \bar{L}  M} \gamma^M_{IJ}$
&
\\
&
$~~~~
-\frac{1}{4} \mathcal{I}^1_{I \bK  A}\d_{  J \bbL} \bar \Gamma^A  \bab^{\bbL} +
\frac{1}{4} \bar{\mathcal{I}}^1_{\bK I  A}\d_{  \bbL J}  \Gamma^A  \bab^{\bbL} $
&   \parbox{0.1\textwidth}{\beq ~\eeq}
\\    \hline
    \end{tabular}
\end{center}
where $\mathcal{I}^1$ and $\tau_{AB}$
 were introduced in eq. \eqref{iato1} and \eqref{eq_tau}, and we also defined
\bea
  \mathcal{I}^2_{IJ\bK\bbL}&\equiv&i\d_\bbL \xi^{1}_{[IJ]\bK} 
 -i\d_\bK  \xi^{1}_{[IJ]\bbL} 
 +i\p_{J} \xi^{2}_{[\bK \bbL]  I} 
 -i \p_{I} \xi^{2}_{[\bK \bbL]  J}
\nl
\mathcal{I}^3_{A I} &\equiv& 
 -\p_I \bg^B \left (\zeta^2_{B \bI}      (e^V T_A e^{-V}\bl)^\bI
 +\zeta^{2\bar{M}}_{\bK \bar{N}}(e^{-V} T_A e^V)_{\bar{M}}^{\bar{N}}     (\bar T_B \bl)^\bK\right) \nl
&& -   \p_I \g^B  \left (\zeta^1_{B K} (T_A\l )^K 
 + \zeta^1_{A K} (T_B\l )^K\right ) ~.
\eea
The Lie derivative of objects containing a global symmetry index 
such as $\li(\bar \zeta^2_{AI})$ is defined in eq.~\eqref{lie_global_symmetry_index}.

\subsubsection{Comments on the consistency conditions}
We have presented the set of consistency conditions for the \sLRG~anomaly.
In the original formulation of the \LRG~equation, 
there are just three consistency condition which 
can not be used to eliminate some anomaly coefficients as an algebraic
function of the others (see eqs.~$(2.66)$ and $(2.68)$ of \cite{Baume:2014rla});
these three equations are
 similar in form to eqs.~\eqref{eq_gradient_flow}, \eqref{eq_vector_beta_function}, and \eqref{eq_beta_eta};
 in the superspace formalism we find many  more consistency conditions.
This is to be expected, as the anomaly discussed here is related to the super-Weyl symmetry, which is a larger symmetry than the one considered in the \LRG~equation. 
Moreover, one has to bare in mind that the \sLRG~anomaly coefficients are not in one-to one correspondence with the ones of the \LRG~ equation.
Another crucial difference is appearance of the unknown functions $\Upsilon$ and $\Omega$, which makes the equations less constraining. 

Finally, we mention that the equations associated with the following terms in $\delta^{WZ}_{\s_1,\s_2}$ are vanishing by imposing the previous constraints, thus providing a highly non-trivial consistency check for our set of equations: \\
$ \commuB R \N_\a \l^i,~~ \commuA  \bar{R} R,~~\commuA   R\N^2\l^i,
~~\commuA   R\N^\a\l^i\N_\a \l^j, ~~\commuA  \bDm^2 \bar{R},$ \\ 
$
\commuA  \N^\a\l^i \N_\a R  , ~~ \commuA  \N^{\a \da} \l^i G_{\a \da}~.
$

 \section{General implications}
\label{sec_implications}
 \subsection{$a$-function and $a$-maximization}
 \label{sec_gradient_flow_and_a_max}
The anomaly coefficient $a$ at the fixed point
is related to the irreversibility of RG flow
(the $a$-theorem) \cite{Cardy:1988cwa,Jack:1990eb,Osborn:1991gm,Komargodski:2011vj,Komargodski:2011xv}
and gives a constraint on the number of degrees
of freedom which can emerge in the IR from a given UV
description.
In the case of supersymmetric theories $a$
can be used to determine the superconformal R-charge
at criticality and the anomalous dimension of chiral primaries
using a-maximization \cite{Intriligator:2003jj}. 
In the context of the local RG equation, the function $\tilde a$
 is a continuation of the anomaly coefficient $a$ off criticality.
 As was shown in \cite{Jack:1990eb,Osborn:1991gm},
away from conformal fixed point the quantity $\tilde{a}$
   is ambiguous up to terms which are quadratic in the beta functions
   (see eq.~(\ref{aaaambiguity})). Moreover, it is shown that 
   at leading order
   $\tilde{a}$ is monotonically decreasing along
 the RG flow;   this feature, if valid non-perturbatively (perhaps just in some specific scheme), would correspond to a strong
  version of the $a$-theorem, in analogy with the two-dimensional case \cite{Zamolodchikov:1986gt}.
  
In this section we will 
study the constraints on $\tilde{a}$ 
 in the superspace formulation of the local RG equation.
 This quantity was already defined in eq.~(\ref{aaatilde})
 and was shown to be real by the consistency condition eq.~(\ref{aaareal}).
  We will also discuss $a$-maximization,
and its generalization off-criticality, proving a conjectured exact formula for this quantity. 

In the superspace formulation, $\tilde{a}$ satisfies
\bea
 \d_I\tilde a&=& \bar \beta^\bJ\chi_{I\bJ}-\beta^J (\d_J\bar w_I - \d_I \bar w_J) \, ,
 \label{eq_gradient_flow3}
\eea
which follows combining eqs.~(\ref{eq_gradient_flow2},\ref{aaareal}).
This equation has a direct analog also in the generic non-supersymmetric case
and gives non-trivial relations between $\beta$-functions (see, e.g. \cite{Jack:2013sha});
it is mostly famous for providing a proof for the irreversibility of the 
RG flows in perturbation theory \cite{Osborn:1991gm,Jack:1990eb}.
Indeed, by multiplying by $\beta$-functions, and using the reality condition \eqref{aaareal} we find
\bea
\label{eq_tilde_a_irrevesibility}
 \mu \frac{d}{d\mu} \tilde a \equiv \left (\beta^I\d_I +\bab^\bJ\d_\bJ\right ) \tilde a &=& \beta^I \bab^\bJ(\chi_{I\bJ} + \bar \chi_{\bJ I})
\eea
At leading order in $\beta$ and $\d\beta$ we can use eqs.~(\ref{zamolo},\ref{positivismo}) to show that 
\bea
\chi_{I\bJ} + \bar \chi_{\bJ I}\propto \mathcal G_{I\bJ}+O(\beta,\d\beta)
\eea
where $\mathcal G$ is the positive definite matrix defined in \ref{zamolo}.
 We conclude that the RHS of \eqref{eq_tilde_a_irrevesibility} is positive definite, 
 and $\tilde a$ is therefore a function of the couplings which
  changes monotonously along the RG flow (at leading order), thus excluding the possibility of 
cyclic flows \cite{Luty:2012ww,Fortin:2012hc,Fortin:2012hn}. 
As was demonstrated in \cite{Baume:2014rla}, there exists a scheme in which 
$\tilde a$ coincides with a quantity related to the on-shell dilaton amplitude, which shares the same property.

The quantity $\Xi_{JI}=\d_J\bar w_I - \d_I \bar w_J$ 
was found to vanish in some appropriate
 scheme\footnote{Due to the transformation law in eq.~\eqref{aaaambiguity},
the relation $\Xi_{JI}=0$ can not be valid in all the schemes.}
 in every example, to all orders checked. If one could prove, in addition to
the positivity of $\chi_{I \bar{J}}$, that $\Xi_{JI}$ vanishes  one would establish
the gradient flow property for $\tilde{a}$ (see eq.~\eqref{eq_gradient_flow3}). 
The consistency condition in eq.~\eqref{eq_rotore} can be used to re-express
$\Xi_{JI}$ in term of $\xi^2$ and $\zeta^2$ anomaly coefficients,
but does not give us any argument supporting  
the conjecture that $\Xi_{JI}=0$ outside the fixed point.

\subsubsection{$a$-maximization}
\label{sec_a_max}

In this section we assume the theory can be formulated in term of
some elementary field description, which corresponds to assuming that
the theory is asymptotically free and can be thought as in the UV as a deformation 
of a free field theory. Also, in this section the lower-case indices $i,j$ are running on the 
elementary field content of the theory.

The idea behind the "$a$-maximization" prescription is the following: 
the superconformal $R$ symmetry is a useful tool
in the study of RG flows, because the $R$ charge of a field determines also its scaling dimension
\bea
\label{eq_R_gamma}
R_{f,i}&=&\frac{2\gamma_i-1}{3} 
\eea
where $R_{f,i}$ is the $R$-charge of the fermion in the superfield $\Phi_i$ and 
$\gamma_i$ is the corresponding  anomalous dimension,
 which is assumed to be diagonalized.
  We can write 
(after setting the couplings to their background values, 
and diagonalizing the matrix $\Gamma^A T_A$):
\bea
\label{eq_R_gamma2}
\gamma_i^j&=&\Gamma^A (T_A)_i^j~,
\eea
where the matrix $ (T_A)_i^j$ is acting on the elementary field content.

If in a theory there are some unbroken not-R $U(1)_A$ symmetries, the 
candidate R-symmetry is ambiguous, because we can generate another R-symmetry
by shifting the original  one by some linear combination of the $U(1)_A$s.
The correct superconformal R-symmetry can be fixed using the
property that at the fixed point the anomaly coefficient of each
$U(1)_A \times {\rm Gravity}^2$ and $U(1)_A \times U(1)_R^2$
are equal\footnote{There is a potential loophole due to possible IR accidental symmetries
which may mix with the $U(1)_R$.}
 \cite{Intriligator:2003jj}; this is equivalent to $a$-extremization, where 
the central charge $a$ is a calculated from the fermionic 
R-charges using \cite{Anselmi:1997am,Anselmi:1997ys}.

Concretely, the following equation which must be satisfied by each of the unbroken $U(1)$ symmetries:
\bea
9 Tr [R_f^2 T_A] - Tr [T_A]&=&0 \, ,
\eea
where again the trace is in the space of the elementary fields.
Writing this equation using \eqref{eq_R_gamma} we find the constraint
\bea
Tr [T_AT_B]\Gamma^B-Tr [ T_AT_BT_C ]\Gamma^B\Gamma^C &=&0~.
\eea
Let us compare this equation with the consistency condition \eqref{amaxgeneralized}, 
evaluated at the fixed point, for a $U(1)$ symmetry which is unbroken by the 
background spurion\footnote{Recall that we are 
considering chiral anomalies with abelian symmetries only, 
therefore all the indices in this equation correspond to $U(1)$ symmetries.}:
 \bea
 \label{eq_a_max_Upsilon}
 2 \kappa_{AB} \g^B 
 +3k_{ABC}
 \Gamma^B\Gamma^C
 &=&  
-\Upsilon_A~. 
\eea
Using eqs.~(\ref{eq_k_ABC},\ref{eq_k_AB}) 
 we find that the $a$-maximization equation 
corresponds to this consistency condition when $\Upsilon_A$ vanishes.
Indeed, under the assumption that 
there are no chiral singlet functions,
$\Upsilon_A$ is a constant and  it can always be set to zero
 using the reparameterization described in section \ref{sec_ambiguity}, see eq.~\eqref{ambi3}.
In an expansion nearby a free fixed point,
the parameterization
for which $\Upsilon_A=0$ coincides with the one usually used in perturbative 
calculation (in which the eigenvalues of $\Gamma$ are the anomalous dimensions of elementary fields).

We stress that eq.~(\ref{amaxgeneralized}) generalizes a-maximization in two ways,
because it is valid also outside the conformal fixed point and for symmetries which
are explicitly broken by some coupling. Setting $\Upsilon_A=0$ in eq.~(\ref{amaxgeneralized}) we get:
\beq
 2 \kappa_{AB} \g^B +3k_{ABC}\Gamma^B\Gamma^C -  \zeta^1_{A I}  \beta^I 
-w_\bI (e^V T_A e^{-V}\bl)^\bI =
 - \Omega _I (T_A\l)^I \, .
 \label{eq_omegone1}
\eeq
$\Omega _I$ is not an anomaly coefficient, but 
some arbitrary function which appears also (and only) in eq.~(\ref{chisolution}):
\beq
\chi_{I \bJ}- \p_I w_\bJ 
-   2i \xi^{1}_{[IK] \bJ}  \b^K  
  +  \zeta^1_{A I} \p_\bJ \g^A
=  - \p_\bJ \Omega_I \, .
 \label{eq_omegone2}
\eeq
This equation can be used 
to relate $\chi$ to other anomaly coefficients
(this is possible just modulo the function $\Omega _I$, which is not directly an anomaly coefficient).
These two equations have no analog in the case without supersymmetry.

\subsubsection{$\tilde a$ off-criticality}
\label{sec_a_tilde}

We can now start with the differential equation for $a$ eq.~\eqref{eq_gradient_flow}
and use eq.~\eqref{chisolution} to eliminate $\chi$ and combine with \eqref{amaxgeneralized}.
We finally find the following relation:
\bea
 \d_\bI \left (\tilde a+ \Upsilon_A\Gamma^A
+\kappa_{AB}\Gamma^A\Gamma^B + k_{ABC}\Gamma^A\Gamma^B\Gamma^C
+\beta^I  \tilde \Omega_I 
\right)&=&0
\label{eq_almost_tilde_a}
\eea
where we introduced the notation
 \bea
 \tilde \Omega_I\equiv \Omega_I-\bar w_I ~.
 \label{notazi-fugazi}
 \eea 
This result implies that the expression in parentheses is a function of the chiral sources, which is invariant under all global symmetries of the theory.  In the absence of chiral singlets, as discussed in section \ref{sec_flavor_symmetry}, such a function must be a constant. Working in the parameterization with vanishing $\Upsilon$, we conclude that $\tilde a$ has the following form
\bea
 \tilde a&=&-\kappa_{AB}\Gamma^A\Gamma^B - k_{ABC}\Gamma^A\Gamma^B\Gamma^C-\beta^I \tilde  \Omega_I +const.
 \eea
Under the same assumption, the chiral functions $\kappa_{AB}$ and $k_{ABC}$ must be constant, and eqs.~(\ref{eq_k_ABC}, \ref{eq_k_AB}) can be used off-criticality. We find the expression for $\tilde a$ in terms of the anomalous dimensions matrices, as advertised in the introduction:
\bea
\tilde a &=&-\frac{1}{128\pi^2} Tr [\gamma^2] + \frac 1 {192\pi^2} Tr[\gamma^3 ]  -\beta^I \tilde  \Omega_I +{\rm const.}
\label{eq_atildata}
\eea   

At the conformal fixed point, $\tilde{a}$ coincides with the central charge $a$ and  
eq.~\eqref{eq_atildata} coincides with the expression found in \cite{Anselmi:1997am,Anselmi:1997ys}.
Out of criticality, eq.~\eqref{eq_atildata} was
conjectured and tested in \cite{Freedman:1998rd,Kutasov:2003ux,Kutasov:2004xu,Barnes:2004jj,Jack:2013sha,Jack:2014pua,Erkal:2010sh} \footnote{In \cite{Kutasov:2003ux,Kutasov:2004xu,Barnes:2004jj} the term proportional to $\beta$
in~\eqref{eq_atildata}
 was introduced as a Lagrange multiplier, whose purpose was to 
 ensure the vanishing of the $\beta$ function at the fixed point. 
}.
Moreover, the same line of argument lead to an interesting conjecture regarding the structure of anomalous dimensions in renormalizable supersymmetric field theories (see eq. (4.17) of \cite{Jack:2014pua} which is an extension of an earlier equation appearing in \cite{Jack:2013sha}). Our equation \eqref{amaxgeneralized} looks like a generalization of this conjectured formula.
The advantage of our method is that these results are derived as consistency conditions, rather than conjectured, and are valid off-criticality by definition.
The downside is that in the presence of chiral singlets, the most we can get is eq. \eqref{eq_almost_tilde_a}.

As a last comment, we recall that any dependence of $\tilde \Omega$ on the $\beta$-function can be removed by a choice of scheme (see discussion in section \ref{sec_scheme}). 

 \subsection{The conformal manifold}
\label{sec_manifold}

When we perturb a conformal field theory with 
a marginal operator, the deformation preserve conformality
just at zero order because in general the deforming operator dimension
itself gets corrections: we can further classify such perturbations
into marginally relevant, marginally irrelevant and exactly marginal.
One may expect that the case of exactly marginal deformations
is fine-tuned in absence of special symmetries; 
indeed the only known examples of such deformations in $d=4$
are realized in supersymmetric theories \cite{Leigh:1995ep, 
Aharony:2002hx,Kol:2002zt,Kol:2010ub,Green:2010da}.

In general one can have multiple exactly marginal deformation
which parameterize a continuous family of conformal field theories.
Assuming that there are no singularities, one can
locally think of this family of CFTs as a submanifold of the 
space of all the couplings of the theory.
Such objects are referred as Conformal Manifolds (CM), 
and appear whenever the number of 
independent conditions, which are necessary
for the vanishing of all the $\b$-functions is less than the 
number of coupling constants in the theory \cite{Leigh:1995ep}.
Nearby a given $\mathcal{N}=1$ SCFT,
 the conformal manifold can be built as the quotient of
all the marginal deformation of the theory divided
by the complexified symmetry group \cite{Green:2010da}.
This explains why the existence of a CM is a rather common property 
in the case of supersymmetric theories.

In this section we will apply the superspace formulation 
of the local RG equation for the study 
of supersymmetric CM. 
Along the CM all the $\b$-functions $\b^I=0$;
also, we can always choose the coordinates in the space of the couplings
in such a way that nearby the origin $\l^I=0$ the conformal manifold directions 
are tangent to some group of coordinates which we denote with $\hat{I}$.
Along these directions 
\bea
\label{eq_vanishing_beta}
 \d_{\hat{I}} \beta^J=\d_{\hat{\bI}} \beta^J = 0 \, .
\eea
Moreover we should identify any two points in $\lambda^I$ space,
 which are related by a global symmetry transformation
\bea
\lambda^I &\sim& \lambda^I + (T_A\l)^I \, .
\eea
Deformation by descendant operators are non-physical as they do not 
modify the action. According to the Ward identities of the flavor symmetries,
 the vectors $(T_A\l)^I$ correspond to operators which are descendants of  currents.  
The CM direction are then characterized by the following property
\bea
\label{eq_vanishing_P}
\d_{\hat{I}}\bar \Gamma^A = \d_{\hat{\bI}} \Gamma^A=0 \, ,
\eea
which is the constraint on primary operators $\OO_{\hat{I}}$ and $\bar \OO_{\hat{\bI}}$ derived in section \ref{sec_superconformal}.
The properties in \eqref{eq_vanishing_beta} and \eqref{eq_vanishing_P} 
have the following interesting implication:
if we consider covariant functions which have legs just in  the conformal manifold $\hat{I},\hat{\bI}$,
 the Lie derivatives $\li, \bli$ defined in section \ref{sec_Lie_derivative} are vanishing.

We can now write the consistency conditions discussed in section \ref{sec_consistency_conditions}, specializing to the conformal manifold (namely, imposing equations  \eqref{eq_vanishing_beta} and \eqref{eq_vanishing_P} for
indices along the CM).
\bit
\item Eqs. \eqref{aaareal}, \eqref{eq_gradient_flow} and \eqref{eq_gradient_flow2} 
we find that $a$ is real and constant along the manifold
\bea
 a - \bar a = \d_{\hat{I}} a = \d_{\hat{\bI}} a = 0 \, .
\eea
\item
Specializing to components tangent to the manifold, 
 eqs.~\eqref{positivismo}  and (\ref{def_metrica_zamolo}) tell us that
\bea
-\frac{\chi_{\hat{I} \hat{\bJ}}}{8} = \epsilon^1_{\hat{I} \hat{\bJ}} = g_{\hat{I} \hat{\bJ}} \, ,
\eea
from which we find that the hermitean part of $\chi_{\hat{I} \hat{\bJ}}$ is proportional to the Zamolodchikov metric
\bea
\chi_{\hat{I} \hat{\bJ}} + \bar \chi_{\hat{\bJ}   \hat{ I}} \propto \mathcal G_{\hat{I}  \hat{\bJ}} \, .
\eea
\item Eq. \eqref{eq_rotore} tells us that along the manifold 
\bea
\d_{\hat{I}} \bar w_{\hat{J}} = \d_{\hat{J}} \bar w_{\hat{I}}
\eea
 and therefore locally we can write 
 \bea
 \bar w_{\hat{J}} =\d_{\hat{J}}  \omega \, .
 \eea
\item Eqs.~(\ref{eq_useful},\ref{chisolution}) give
\bea
\bar \chi_{{\hat{\bJ}} {\hat{I}}} - \d_{\hat{\bJ}} \bar{w}_{\hat{I}} = 
\chi_{{\hat{I}} {\hat{\bJ}} }-\d_{\hat{I}}  w_{\hat{\bJ}} = -\d_{\hat{\bJ}} \Omega_{\hat{I}} \, ,
\eea
therefore $\Omega_{\hat{I}}$ is constrained by the relation
 $\d_{\hat{\bJ}} \Omega_{\hat{I}} =\d_{\hat{I}} \bar{\Omega}_{\hat{\bJ}}$~.
We finally find 
\bea
\mathcal G_{ {\hat{I}} {\hat{\bJ}}} \propto \chi_{ {\hat{I}} {\hat{\bJ}}} + \bar \chi_{{\hat{\bJ}} {\hat{I}}}
 = \d_{\hat{\bJ}} (-2 \Omega_{\hat{I}}  + \d_{\hat{I}} (\omega +\bar \omega))
\eea
and conclude that the Zamolodchikov metric on the conformal manifold is K\"ahler,
because it satisfies
\beq
\d_{\hat{K}} \mathcal{G}_{ {\hat{I}} {\hat{\bJ}}}-\d_{\hat{I}} \mathcal{G}_{{\hat{K}} {\hat{\bJ}}}=0 \, , \qquad
\d_{\hat{\bK}} \mathcal{G}_{{\hat{I}} {\hat{\bJ}}}-\d_{\hat{\bJ}} \mathcal{G}_{ {\hat{I}} {\hat{\bK}}} =0 \, .
\eeq
An earlier proof of this result was found in \cite{Asnin:2009xx}.
\eit

 \section{Conclusions}
\label{sec_conclusions}

In this work we presented the superspace formulation of the local RG equation, and derived the Wess-Zumino consistency conditions associated with it. We found that many of the results in the literature regarding supersymmetric RG flows, which were derived using a variety of methods, appear naturally within this framework. However, we believe that there is much room for further exploration. Here are just some of the ideas for future research directions: 
\bit
\item We found physical interpretation for only a few of the consistency conditions.
The choice of basis for the consistency conditions given here is not necessarily the optimal one, and perhaps there are alternative representations of the equations which makes the physical meaning more visible.

\item Our work here was restricted to theories with chiral anomalies involving only $U(1)$ symmetries.
The solution we found for the consistency condition relating the chiral  and Weyl anomalies proved instrumental in the derivation of the 
results of section \ref{sec_gradient_flow_and_a_max}, but perhaps a treatment of non-abelian global anomalies may uncover more interesting structure.

\item The curved background was defined using the old-minimal formulation of supergravity. This required the usage of the chiral compensator and the Ferrara-Zumino multiplet. However, in order to have a better understanding of the constraints on RG flow related to $R$-symmetry, it could prove useful to use the so called $\mathcal R$ multiplet, whose lowest component is the $R$-current 
(see e.g.~\cite{Gates:1983nr}).
 Examples for the usage of this multiplet are given in \cite{Abel:2011wv,Buican:2011ty}. This adjustment might be non-trivial, as it requires switching to the new-minimal formulation of supergravity, in which some of the identities used in our computation are no longer valid.

\item Another possible extension of this work is to consider theories with extended supersymmetry, and to check whether the consistency conditions reveal more structure in the RG flow of such theories.

\item It would be interesting to extend this analysis in different dimensions.
The case of $d=6$, where an $a$-theorem is still lacking  \cite{Elvang:2012st,Grinstein:2013cka,Grinstein:2014xba,Osborn:2015rna}, 
is especially interesting; it could be that supersymmetry adds essential ingredients. 

\eit

\section*{Acknowledgments}

We are grateful  to R. Rattazzi for many useful comments and to H. Osborn for the hospitality and helpful discussions.
We would also like to thank L. Bonora for a useful discussion.
This work was supported by the Swiss National Science Foundation under grant 200020-150060.

\section*{Appendix}
\addtocontents{toc}{\protect\setcounter{tocdepth}{1}}
\appendix

\section{Formulas and conventions}
 \subsection{Classical SuperWeyl variations}
The calculations are done in superspace in old minimal supergravity;
 the notations and the conventions of \cite{Buchbinder:1995uq} are used.
 Other useful references are: \cite{Gates:1983nr,Wess:1992cp,Howe:1978km,Kaplunovsky:1994fg,Schwimmer:2010za,Grosse:2007au}.
 Super-Weyl transformation will be parameterized by the chiral superfield $\s$.
We denote with $E$ the determinant of the supervielbein
and with $\varphi$ the conformal compensator; the gravitational superfields
background is described in terms of the chiral fields  $W_{\a \b \gamma}$ and $R$
(whose components contain the Weyl tensor and the scalar curvature)
and of the real $G_{\a \da}$ (which contains the traceless Ricci tensor).
Under a super Weyl transformations, these fields transform as:
\begin{center}
\begin{minipage}
{0.7\textwidth}
\begin{tabular}{  >{$}l<{$}   >{$}l<{$}     >{$}l<{$}  }
\delta_\sigma \varphi &=&\sigma \varphi\nl
\delta_\sigma E^{-1}& =& (\sigma+\bar{\sigma}) E^{-1}\nl
\delta_\sigma R    &=& (\bar{\sigma} -2 \sigma) R -\frac1 4 \bar{\mathcal{D}}^2 \bar{\sigma}\nl
\delta_\sigma G_{\alpha \dot{\alpha}}&=& -\frac1 2 (\sigma+\bar{\sigma}) G_{\alpha \dot{\alpha}} -  \mathcal{D}_{\alpha \dot{\alpha}} (i \sigma-i \bar{\sigma})\nl
\delta_\sigma W_{\alpha \beta \gamma}&=&-\frac{3}{2} \sigma W_{\alpha \beta \gamma}\\
\end{tabular}
\end{minipage}
\begin{minipage}[b]{0.2\textwidth}
\beq ~\eeq
\end{minipage}
\end{center}
The Weyl variation of SUSY derivatives (specializing to derivatives acting on scalars):
\begin{center}
\begin{minipage}
{0.7\textwidth}
\begin{tabular}{  >{$}l<{$}   >{$}l<{$}     >{$}l<{$}  }
\delta_\sigma \Dm_\alpha &=& \half (\sigma -2\bar{\sigma}) \Dm_\alpha\nl
\delta_\sigma \Dm_{\alpha \dot{\alpha}} &=& -\frac{1}{2}(\s+\bs)  \Dm_{\alpha \dot{\alpha}} -\frac{i}{2} ( (\Dm_\alpha \sigma) \bar{\Dm}_{\dot{\alpha}}  + (\bar{\Dm}_{\dot{\alpha}} \bar{\sigma}) \Dm_\alpha)\nl
\delta_\sigma \Dm^2& =& (\sigma-2 \bar{\sigma}) \Dm^2 +2(\Dm^\alpha \sigma) \Dm_\alpha\nl
\end{tabular}
\end{minipage}
\begin{minipage}[b]{0.2\textwidth}
\beq ~\eeq
\end{minipage}
\end{center}
where $\Dm_{\alpha \dot{\alpha}} = \frac i 2 \{\Dm_{\alpha},\bar \Dm_{ \dot{\alpha}}\}$.

\subsection{Covariant derivatives}
Covariant derivatives acting on chiral fields
\begin{center}
\begin{minipage}
{0.7\textwidth}
\begin{tabular}{  >{$}l<{$}   >{$}l<{$}     >{$}l<{$}  }
A^{\a I}_J &=& (e^{-V})^{I\bar K} (\mathcal D^\a e^V)_{\bar K J}
\nl
\N^\a e^V &=&0\nl
\N^\a \l^I &=&\mathcal D^\a\l^I + (A^\a)_J^I\l^J
\nl
\N^{\a \da} \l^I &= &\Dm^{\a \da} \l^I +\frac{i}{2} \bDm^\da A^{\a I}_J \l^J
\nl
\N^2 \l^I &=&D^2 \l^I +2 A^{\a I}_J D_\a \l^J +(A^{\a I}_J A^J_{\a K} +\Dm^\a A^I_{\a K} ) \l^K \nl
\end{tabular}
\end{minipage}
\begin{minipage}[b]{0.2\textwidth}
\beq ~\eeq
\end{minipage}
\end{center}
Covariant derivatives acting on anti-chiral fields
\begin{center}
\begin{minipage}
{0.7\textwidth}
\begin{tabular}{  >{$}l<{$}   >{$}l<{$}     >{$}l<{$}  }
B^{\da \bI}_\bJ &=& (\bar {\mathcal D}^\da e^{V})_{\bar J K} ( e^{-V})^{ K \bar L}
\nl
\bN^\da e^V &=&0\nl
\bN^\da \bl^\bI &=&\bar {\mathcal D}^\da\bl^\bI + (B^\da)_\bJ^\bI\bl^\bJ
\nl
\N^{\a \da} \bl^\bI &= &\Dm^{\a \da} \l^I +\frac{i}{2} \Dm^\a B^{\da \bI}_\bJ \bl^\bJ
\nl
\bN^2 \bl^I &=&\bar D^2 \bl^\bI  + 2 B^{\bI}_{\da \bJ} \bar D^\da \bl^\bJ 
+(B^{\bI}_{\da \bJ} B^{\da \bJ}_{\bK} +\bar {\Dm}_\da B^{\da \bI}_{\bK} ) \bl^\bK \nl
\end{tabular}
\end{minipage}
\begin{minipage}[b]{0.2\textwidth}
\beq ~\eeq
\end{minipage}
\end{center}

 \subsection{Useful identities}
\begin{center}
\begin{minipage}
{0.7\textwidth}
\begin{tabular}{  >{$}l<{$}   >{$}l<{$}     >{$}l<{$}  }
\bDm^\da G_{\a \da}&=&\Dm_\a R \nl
\bN_\da W_\a^{\pm} &=&0\nl
\N^\a W_\a^{\pm} &=& \bN_\da (\bar{W}^\pm)^\da\nl
\N_\a \N^2 \l^I&=&4 \bar{R} \N_\a \l^I \nl
\bN^2 \N_\a \l^I &=&4 R \N_\a \l^I -4 (W_\a  \l)^I \nl
\bN_\da \N^2 \l^I &=& 4 (G_{\a \da}+i \N_{\a \da}) \N^\a \l^I -4 (\bar{W}^+_\da \l)^I\nl
\N^\a \bN_\da \N_\a \l^I& =& -\frac{1}{2} \bN_\da \N^2 \l^I -2 G_{\a \da} \N^\a \l^I +2 (\bar{W}_\da^+ \l)^I\nl
\bN^2 \N^2 \l^I &=& - 8i G_{\a \da} \N^{\a \da} \l^I   -8 \N_{\a \da} \N^{\a \da} \l^I\nl
&& +4 \Dm^\a R \N_\a \l^I + 8 R \N^2 \l^I -8 (W^\a)^I_K \N_\a \l^K -4 (\N^\a W_\a)^I_K \l^K
\end{tabular}
\end{minipage}
\begin{minipage}[b]{0.2\textwidth}
\beq ~\eeq
\end{minipage}
\end{center}
where we used the notations
\begin{center}
\begin{minipage}
{0.7\textwidth}
\begin{tabular}{  >{$}l<{$}   >{$}l<{$}     >{$}l<{$}  }
(W_\a^+)^I_J \equiv (W_\a)^I_J & \qquad \qquad &(\bar{W}_\da^+)^I_J \equiv(e^{-V})^{I \bar L} (\bar{W}_\da)^\bK_{\bar L} (e^V)_{\bK J} \, ,
\nl
(\bar{W}_\da^-)^\bI_\bJ \equiv (\bar{W}_\da)^\bI_\bJ
&\qquad\qquad &
(W_\a^-)^\bI_\bJ \equiv  (e^V)_{\bJ K}  (W_\a)^K_L (e^{-V})^{L \bI} \, .
\end{tabular}
\end{minipage}
\begin{minipage}[b]{0.2\textwidth}
\beq ~\eeq
\end{minipage}
\end{center}

\section{Useful definitions and results}
\subsection{Notations}
\begin{center}
\begin{minipage}
{0.7\textwidth}
\begin{tabular}{  >{$}l<{$}   >{$}l<{$}     >{$}l<{$}  }
\beta^I &\equiv& b^I + \Gamma^A (T_A\l)^I\nl
U_I^J&\equiv&\delta_I^J +\d_I\beta^J\nl
\Lambda^I &\equiv& (U^{-1})^I_K (\N^2 \l^K +4 \beta^K \bar{R}) \nl
\Pi^{IJ}&\equiv&\N^\a \l^I \N_\a \l^J -\frac{1}{2} (\b^I \Lambda^J+\beta^J \Lambda^I) 
\end{tabular}
\end{minipage}
\begin{minipage}[b]{0.2\textwidth}
\beq ~\eeq
\end{minipage}
\end{center}

\subsection{Lie derivatives in parameter space}
The Lie derivatives of a covariant function $Y_{IJ\ldots}^{K\ldots}$, where the indices $I,J,\ldots$ run over the marginal operators, is given by
\bea
\li(Y_{I\bJ \ldots}^{K\ldots})
&\equiv&\beta^L\d_LY_{I\bJ \ldots}^{K\ldots}
+\gamma_I^LY_{L\bJ \ldots}^{K\ldots}
+\bar {\tilde \gamma}_{\bJ}^{\bar L}Y_{I\bar L \ldots}^{K\ldots}
+\ldots
-\gamma_L^KY_{I\bJ\ldots}^{L\ldots} 
- \ldots\nl
\bli(Y_{I\bJ \ldots}^{K\ldots})
&\equiv&\bar \beta^{\bar L}\d_{\bar L}Y_{I\bJ \ldots}^{K\ldots}
+\tilde \gamma_I^LY_{L\bJ \ldots}^{K\ldots}
+\bar \gamma_{\bJ}^{\bar L}Y_{I\bar L \ldots}^{K\ldots}
+\ldots
-\tilde  \gamma_L^KY_{I\bJ\ldots}^{L\ldots} 
- \ldots
\eea
For tensors which multiply $W_\a^A$ we find
\bea
\li(Y_{A\ldots}^{\ldots})
&\equiv&\beta^I\d_IY_{A\ldots}^{K\ldots}
+(\gamma_A^B+if^B_{AC}\Gamma^C)Y_{B\ldots}^{\ldots}
+\ldots
\nl
\bli(Y_{A\ldots}^{\ldots})
&\equiv&\bar \beta^{\bar I}\d_{\bar I}Y_{A\ldots}^{\ldots}
+ \gamma_A^BY_{ B\ldots}^{\ldots}
+\ldots
\label{lie_global_symmetry_index}
\eea
For tensors which multiply $\bar W_\da^A$ we find
\bea
\li(Y_{A\ldots}^{\ldots})
&\equiv&\beta^I\d_IY_{A\ldots}^{K\ldots}
+\bar \gamma_A^BY_{B\ldots}^{\ldots}
+\ldots
\nl
\bli(Y_{A\ldots}^{\ldots})
&\equiv&\bar \beta^{\bar I}\d_{\bar I}Y_{A\ldots}^{\ldots}
+(\bar \gamma_A^B+if^B_{AC}\Gamma^C)Y_{ B\ldots}^{\ldots}
+\ldots
\eea
The anomalous dimension matrices are given by
\begin{center}
\begin{minipage}
{0.7\textwidth}
\begin{tabular}{  >{$}l<{$}   >{$}l<{$}     >{$}l<{$}  }
\gamma_I^J&\equiv&\d_I\beta^J\nl
\wt  \gamma_I^J&\equiv&\d_I \Gamma^A (T_A \l)^J\nl
\gamma_A^B
&=&\d_I \Gamma^B (T_A \l)^I\nl
\end{tabular}
\end{minipage}
\begin{minipage}[b]{0.2\textwidth}
\beq ~\eeq
\end{minipage}
\end{center}
It is also useful to introduce the following notations
\begin{center}
\begin{minipage}
{0.7\textwidth}
\begin{tabular}{  >{$}l<{$}   >{$}l<{$}     >{$}l<{$}  }
\gamma^I_{JK} &=& (U^{-1})^I_L \p_{KJ} \b^L
\nl
\gamma^{IJ}_{KL} &=& \gamma^{(I}_{KL}\beta^{J)}
\end{tabular}
\end{minipage}
\begin{minipage}[b]{0.2\textwidth}
\beq \label{eq_anomalous_dimension_tensors} ~\eeq
\end{minipage}
\end{center}

Using the covariance of the function $Y$ and the consistency condition \eqref{eq_BdotP} one can show that this Lie derivative has the following properties:
\begin{center}
\begin{minipage}
{0.7\textwidth}
\begin{tabular}{  >{$}l<{$}   >{$}l<{$}     >{$}l<{$}  }
\beta^I \li (Y_{I\ldots})&=& \li (\beta^I Y_{I\ldots})\nl
(T_A\l)^I \li (Y_{I\ldots})&=& \li( (T_A \l)^I Y_{I\ldots})\nl
 \end{tabular}
\end{minipage}
\begin{minipage}[b]{0.2\textwidth}
\beq ~\eeq
\end{minipage}
\end{center}
\begin{center}
\begin{minipage}
{0.7\textwidth}
\begin{tabular}{  >{$}l<{$}   >{$}l<{$}     >{$}l<{$}  }
\left [ \li,\bli \right]
&=& 0\nl
\left [ \li,\d_I  \right] Y_{J\ldots}
&=& 
-\d_{IJ}\beta^K Y_{K\ldots}+\ldots\nl
\left [ \li,\d_\bI  \right] Y_{J\ldots}
&=&   -\d_{\bI J}\Gamma^A (T_A\l)^K Y_{K\ldots}+\ldots\nl
\left [ \li,\d_I \right] Y_{\bJ\ldots}
&=&  
   -\d_{I\bJ}\Gamma^A(\bar T_A\bl)^\bbL Y_{\bbL \ldots}+\ldots\nl
\left [ \li,\d_\bI  \right] Y_{\bJ\ldots}
&=& 
-\d_{\bI\bJ}\Gamma^A (\bar T_A\bl)^\bbL Y_{\bbL \ldots}+\ldots\nl
 \end{tabular}
\end{minipage}
\begin{minipage}[b]{0.2\textwidth}
\beq ~\eeq
\end{minipage}
\end{center}
\begin{center}
\begin{minipage}
{0.7\textwidth}
\begin{tabular}{  >{$}l<{$}   >{$}l<{$}     >{$}l<{$}  }
 \li (Y_I) - Y_K \gamma^K_{IJ}\beta^J&=& U_I^J \li (Y_K (U^{-1})^K_J)\nl
 \bli (Y_I) - Y_K \tilde \gamma^K_{IJ}\beta^J&=& U_I^J \bli (Y_K (U^{-1})^K_J)
 \end{tabular}
\end{minipage}
\begin{minipage}[b]{0.2\textwidth}
\beq ~\eeq
\end{minipage}
\end{center}

\subsection{Super-Weyl variations of functions}
\begin{center}
\begin{minipage}
{0.9\textwidth}
\begin{tabular}{  >{$}l<{$}   >{$}l<{$}     >{$}l<{$}  >{$}l<{$}  }
\Delta_\s^\sW ((e^V)_{\bI J}) &=& (\sigma+\bs) \Gamma^A( e^V T_A)_{\bI J} \nl
\Delta_\s^\sW ((A_\a)_J^I) &=& \half (\s -2\bs) (A_\a)_J^I &+~\Dm_\a\s \Gamma_J^I~~~+(\s+\bs)\d_K \Gamma^I_J \N_\a\l^J \nl
\Delta^{\sW}_\s (Y_I \N^\a \l^I)
&=&\half (\s -2\bs) Y_I\N^\a \l^I 
&+~ \Dm^\a \s ~Y_I \b^I \nl
&&+~ \s \li (Y_I)  \N^\a \l^I
&+~\bs   \bli (Y_I)  \N^\a \l^I \nl
\Delta^{\sW}_\s(Y_\bI \bN^\da \bl^\bI)&=& 
\half (\bs -2\s) Y_\bI \bN^\da \bl^\bI 
&+ ~
 \bDm^\da \bs ~Y_\bK \bab^\bK\nl
 &&  + ~\s \li (Y_\bI)  \bN^\da \bl^\bI
&+~\bs   \bli (Y_\bI)  \bN^\da \bl^\bI\nl
\Delta_\s^{\sW}(Y_I\N^2 \l^I)&=&
( \s-2\bs ) Y_I \N^2 \l^I 
&+~  \Dm^2 \s ~Y_K \b^K ~~+ 2 \Dm^\a \s   ~Y_K U^K_I \N_\a \l^I \nl
&&+~ \s \mathcal{L}(Y_I) \N^2 \l^I
&+~\s Y_K\p_{IJ } \b^K \N^\a \l^I \N_\a \l^J \nl
&&+~\bs\bar{\mathcal{L}}(Y_I)  \N^2 \l^I 
&+~ \bs   Y_K \p_{IJ} ( \Gamma^K_L \l^L   )  \N^\a \l^I \N_\a \l^J 
\nl
\Delta^{\sW}_\sigma (Y_I\Lambda^I)
&=& (\s-2\bs)Y_I\Lambda^I 
&+  ~2 Y_I  (\Dm^\a \s) \N_\a \l^I 
\nl
&&+~ \s \li(Y_I) \Lambda^I&
+~\bs \bli(Y_I)  \Lambda^I\nl
&&+~\sigma Y_I \gamma^I_{KL}\Pi^{KL}
&+~\bs Y_I {\gamma}^I_{KL}\Pi^{KL} \nl
\Delta^{\sW}_\sigma (Y_{IJ} \Pi^{IJ})
&=&  (\s-2\bs)Y_{IJ}\Pi^{IJ} &
+~\s (\li(Y_{IJ})- Y_{KL}\gamma^{KL}_{IJ} ) \Pi^{IJ}\nl
&&&+~\bs \le \bar{\mathcal{L}}(Y_{IJ}) -Y_{KL}  {\gamma}^{KL}_{IJ} \ri  \Pi^{IJ} \nl
\Delta_\s^\sW (Y_I \N^{\a\da} \l^I)&=& -\half (\s+\bs)Y_I \N^{\a\da} \l^I &+ ~\Dm^{\a \da} \s Y_I\beta^I\nl
&&-  \frac{i}{2}  Y_I  \d_\bJ\beta^I  \, \Dm^\a \s \bN^\da \bl^\bJ 
&+~\frac{i}{2} Y_I (-\delta^I_J  + \d_J  \Gamma ^I_K \lambda^K)  \bDm^\da \bs \N^\a \l^J \nl
&&+~\s \li(Y_I) \N^{\a\da} \l^I 
&-~ \frac i 2 \s  Y_I \d_{J\bK}\beta^I   \N^\a \l^J  \bN^\da \bl^\bK\nl   
&&+~\bs \bli(Y_I) \N^{\a\da} \l^I 
&-~ \frac i 2 \bs  Y_I \d_{J\bK}\Gamma^I_L \l^L    \N^\a \l^J  \bN^\da \bl^\bK\nl   
\Delta_\s^\sW (Y_A W^{\a A})&=& 
-\frac{3}{2} \s Y_A W^{\a A}&\nl
&&+~i Y_A \Dm^{\a \da} \s \p_\bI \Gamma^A  \bN_\da \bl^\bI
&-~\frac{1}{4} Y_A \Dm^\a \s  \left( \p_\bI \g^A  \bN^2 \bl^\bI + \p_{\bI \bJ} \g^A \bN_\da \bl^\bI \bN^\da \bl^\bJ \right)\nl
&&-~\frac{1}{4}  Y_A  \bDm^2 \bs \d_I\Gamma^A \N^\a \l^i
&+ ~Y_A \bDm_\da \bs   \left(  i \d_I \Gamma^A \N^{\a \da} \l^I  
 +\half\d_{I\bJ}\Gamma^A  \N^\a \l^I  \bN^\da \bl^\bJ  \right)\nl
&& + (\s+\bs) \Big( \li (Y_A)W^{\a A}& +~ Y_A \p_{ I \bJ} \g^A ( -\frac{1}{4}\N^\a \l^I  \bN^2 \bl^\bJ + i  \N^{\a \da} \l^I \bN_\da \bl^\bJ )\nl
&&     &-~\frac{1}{4} Y_A \p_{I\bJ\bK}  \g^A  \N^\a \l^I \bN_\da \bl^\bJ \bN^\da \bl^\bK  \Big)\nl
\end{tabular}
\end{minipage}
\begin{minipage}[b]{0.05\textwidth}
 \beq~\eeq
\end{minipage}
\end{center}

where we used the constraint that $\Gamma$ is hermitean in the sense
of eq.~\eqref{eq_non_renormalization}: $\bar{\Gamma}  e^V=  e^V \Gamma $.

\subsection{Integrations by parts}

The following identities (valid up to total derivatives) are useful in order to perform
some integration by parts which are very common in the task of writing the consistency conditions ($A$ is an arbitrary covariant function of the sources).
\begin{center}
\begin{minipage}
{0.7\textwidth}
\begin{tabular}{  >{$}l<{$}   >{$}l<{$}     >{$}l<{$}  }
\s_{[1} \Dm^2 \s_{2]}  A &=& - \s_{[1} \Dm^\a \s_{2]} \Dm_\a A \nl
\s_{[1} \bDm^2 \bs_{2]} A &=&  \s_{[1} \bs_{2]} \bDm^2 A \nl
\s_{[1} \Dm^{\a \da} \s_{2]}  A_{\a \da}&=&\frac{i}{2} \s_{[1} \Dm^\a \s_{2]} \bDm^\da A_{\a \da} \nl
\s_{[1} \Dm^{\a \da} \bs_{2]}  A_{\a \da} &=&
-\frac{i}{2} \Dm^\a \s_{[1} \bDm^\da \bs_{2]} A_{\a \da}-\frac{i}{2} \s_{[1} \bs_{2]} \bDm^\da \Dm^\a A_{\a \da} \nl
\Dm^\a \s_{[1} \Dm_{\a \da} \s_{2]} A^\da &=&
  \s_{[1} \Dm^\a \s_{2]}  (\frac{i}{2} \bDm_\da \Dm_\a A^\da+\frac{i}{4}  \Dm_\a   \bDm_\da A^\da +i G_{\a \da} A^\da) \nl
  \Dm^\a  \s_{[1} \Dm_{\a \da} \bs_{2]} A^\da &=& \s_{[1}  \bs_{2]} (2 i \bDm_\da (\bar{R} A^\da )-\frac{i}{2} \bDm_\da \Dm^2 A^\da)
+  \Dm^\a \s_{[1} \bDm^\da \bs_{2]} \frac{i}{2} \Dm_\a A_\da\nl
\Dm^\a \s_{[1} \bDm^2 \bs_{2]} A_\a&=&\s_{[1} \bs_{2]} (\Dm^\a \bDm^2 A_\a -\Dm^\a (4 R A_\a))
+\Dm^\a \s_{[1} \bDm^\da \bs_{2]} 2 \bDm_\da A_\a 
\end{tabular}
\end{minipage}
\begin{minipage}[b]{0.2\textwidth}
\beq ~\eeq
\end{minipage}
\end{center}

\subsection{Computation of the consistency conditions}
As an example for the type of computations necessary for deriving the Wess-Zumino consistency conditions, 
we compute the contribution of the $\chi_{I\bJ}$ anomaly.
\bea
\Delta_{\s_2}^{\sW} \le \half E^{-1} \s_1 \chi_{I \bJ} \N_\a \l^I \bN_\da \bl^\bJ G^{\a \da}   \ri 
&=&\half E^{-1}\Big(  \s_1 \chi_{I \bJ} \Dm^{\a \da} (i \bs_2 -i \s_2)  \N_\a \l^I \bN_\da \bl^\bJ
\nl
&&  ~~~~~~~+  \beta^I \chi_{I \bJ}  \Dm_\a \s_2  \bN_\da \bl^\bJ G^{\a \da} 
  -  \beta^\bJ \chi_{I \bJ}    \Dm_\da \bs_2  \N_\a \l^I   G^{\a \da} \nl
&&
  ~~~~~~~+\bs_2 \bli(\chi_{I \bJ} )  \N_\a \l^I \bN_\da \bl^\bJ G^{\a \da}  +\s_2 ( \dots )\Big)~~~~~~~~~~~~~~~~~~~
\eea
The contribution to the consistency condition (see eq. \eqref{eq_delta_WZ}) is 
{\small
\bea
\delta^{WZ}_{\s_1,\s_2} &\supset& \commuA \Bigg(  \le \bli(\chi_{I \bJ} ) -\p_\bJ (\bab^\bk \chi_{I \bK})-\chi_{I \bJ}  \ri
 \N_\a \l^I \bN_\da \bl^\bJ G^{\a \da} 
 -2 i \bab^\bK  \chi_{I \bK}  G^{\a \da}  \N_{\a \da}  \l^I
  -\bab^\bk \chi_{I \bK} \Dm^\a R \N_\a \l^I
\nl
&&\qquad~~~~~~~-  \half \chi_{I \bJ} \N^2 \l^I \bN^2 \bl^\bJ  
- 2\chi_{I \bJ} \N^{\a \da} \l^I \N_{\a \da} \bl^\bJ 
   -\half \chi_{I \bJ} \bN_\da \N^2 \l^I \bN^\da \bl^\bJ -\frac 14\chi_{I \bJ} \N^\a \l^I \N_\a \bN^2\bl^\bJ
   \nl
   &&\qquad~~~~~~~
   + \chi_{I \bJ}  \N^\a \l^I (W_\a^-)^\bj_\bK \bl^\bK 
-\half \p_\bk \chi_{I \bJ} \bN_\da \bl^\bK \bN^\da \bl^\bJ \N^2  \l^I   
   +i \p_\bK  \chi_{I \bJ} \N_\a \l^I \bN_\da \bl^\bK  \N^{\a \da} \bl^\bJ 
\nl
&&\qquad~~~~~~~
-\half \p_K \chi_{I \bJ}  \N^\a \l^K \N_\a \l^I \bN^2  \bl^\bJ
   -i(\p_K \chi_{I \bJ}+  \p_I \chi_{K \bJ}  ) \N_\a \l^I \bN_\da \bl^\bJ \N^{\a \da} \l^K
   \nl
   &&\qquad~~~~~~~
 -\half \p_{K \bbL} \chi_{I \bJ}  \bN_\da \bl^\bbL \bN^\da \bl^\bJ \N^\a \l^K \N_\a \l^I )  
  \nl
  &&+\commuB \le  \beta^I \chi_{I \bJ}   \bN^\da \bl^\bJ G_{\a \da} 
   +\half \chi_{I \bJ} \N_\a \l^I \bN^2 \bl^\bJ 
   + i \chi_{I \bJ}  \N_{\a \da} \l^I \bN^\da \bl^\bJ  
   +\half  \p_\bK \chi_{I \bJ}  \N_\a \l^I  \bN_\da \bl^\bK  \bN^\da \bl^\bJ  \ri
\nl
&&+ \commuC \le \frac1 2 \chi_{I \bJ} \Dm_\a \l^I \bDm_\da \bl^\bJ \ri
\eea
}
\section{Basic examples}

\subsection{The Wess-Zumino model}
\label{sec_WZmodel}
In the Wess-Zumino model each classically marginal coupling $\l^I$ is a trilinear in the 
elementary chiral superfields $\Phi_i$, which are in the fundamental representation
of the $U(N)$ symmetry group. Each coupling index $I$ corresponds to 
a symmetric 3-tensor  product of fundamentals, e.g.: $I=(ijk)$.  
 one finds that the holomorphic $\beta$-function vanishes 
 to all orders\footnote{If one is interested just in composite
 marginal operators, one may use the ambiguity discussed
 in section \ref{sec_ambiguity} to shift $\beta$ and $\gamma$.
  We use in the following discussion the canonical choice $b^I=0$.}
\beq
b^I=0 \, ,
\eeq
and the anomalous dimension matrix $\Gamma^A(T_A)_j^i=\Gamma^i_j$ can be computed perturbatively; 
 the leading and next-to-leading contributions are (see, e.g. \cite{Jack:2013sha}):
\bea
 (\Gamma^{i}_{j})_1&=& \frac{1}{2 (2 \pi)^2} \l^{ikl} \bl^{\bi \bk \bbl} (e^V)_{\bk k}  (e^V)_{\bbl l}   (e^{V})_{\bi j} \nl
 (\Gamma^{i}_{j})_2&=& 
  - \frac{1}{2 (4 \pi)^4} \le (e^V )_{\bj j}   (e^V)_{\bar{k} k}  (e^V)_{\bar{q} r}    \bl^{\bj \bk \bar{q}} \ri
   \left( \l^{rst} \bl^{\bar{r} \bar{s} \bar{t}}  (e^V)_{\bar{s} s}   (e^V)_{\bar{t} t}  (e^V)_{\bar{t} u} \right) \l^{u k i} \, .
 \eea
Using the results of section \ref{sec_components} we read the corresponding terms in the LRG equation
\bea
\beta^I&= \g^I_J \l^J ~~ \qquad &\bab^{\bar{I}}=(\beta^I)^*  \nl
\rho_I^A&= - \p_I \g^A  \qquad  &\rho_{\bar{I}}^A=  \p_{\bar{I}} \g^A \, ,
\eea
where
\bea
\Gamma^I_J=\Gamma^A(T_A)_J^I=\Gamma^{i}_{j} \delta^k_m \delta^l_n + \delta^i_j \g^k_m \delta^l_n +
\delta^i_j \delta^k_m  \g^l_n ~.
\eea

\subsection{Gauge theories}
\label{sec_NSVZ}

In the case of a gauge theory with a simple gauge group, 
the global symmetry group is the product $\mathcal{G}=\Pi_i U(N_i)$,
 where the index $i$ runs on the representations of the matter
fields and each  $U(N_i)$ corresponds to a
group of $N_i$ fields in the same gauge representation.
The symmetry group $\mathcal{G}$ is 
in general broken by the chiral coupling spurions. 

The gauge coupling itself 
can be promoted to a chiral superfield: $S=\frac{4 \pi}{g_h^2} - i \frac{\Theta}{2 \pi} $.
We call $U(1)_K$ the symmetry which  rotates with the same phase all the matter fields 
charged under the gauge group.
Under $U(1)_K$ the superfield $S$ 
transform in a way specified by Konishi anomaly \cite{Clark:1979te,Konishi:1983hf}:
\bea
S \arr S + \frac{ i \Lambda}{ \pi}  \sum_i \, N_i t(r_i)
\eea
 where the index $i$ run on the matter fields gauge representation
  and $t(r_i)$ is the its Dynkin index.
 For our purpose it is convenient to consider the combination $\l_G=e^{-S}$ which 
transforms linearly under  $U(1)_K$
  and  vanishes in the limit of the free theory :
\beq
\l_G\arr \l_G \, e^{ -\frac{ i \Lambda}{ \pi}  \sum_i N_i t(r_i) }  \, .
\eeq
The gauge coupling $\l_G$ does not tranform
under all the other global symmetries in $\mathcal{G}$ which are orthogonal to $U(1)_K$.

Inside each $U(N_i)$ factor, one can diagonalize the
matrix $\Gamma_i=\Gamma^{A_i} T_{A_i}$, where 
the label ${A_i}$ runs on the $U(N_i)$ generators.
We denote with $\gamma^{(a,i)}$ the $N_i$ eigenvalues of the matrix 
$\Gamma_i$,  with $a=1 \dots N_i$;
$\gamma^{(a,i)}$ are the anomalous dimension of the elementary fields
in the gauge representation $i$. 

The action of the global symmetry background gauge field
 on the chiral coupling $\lambda^G$  can be written as
\bea
\Gamma^{A_i}(T_{A_i}\l^G)&=& 
-\frac{1}{\pi}t(r_i) \sum_a \gamma^{(a,i)} \l^G \, .
\eea
The holomorphic $\beta$-function for $\lambda_G$ is given by the one-loop result:
\bea
 b^{\l_G}&=&- \left (\frac{(3t(G)-\sum_i N_i t(r_i))}{2 \pi}\right)\l_G
\eea
This $\beta$-function $ b^{\l_G}$ can be non-zero at
the conformal fixed points.
Using eq.~\eqref{eq_def_beta}, we find the physical $\beta$-function:
\bea
 \beta^{\l_G}&=&-\frac{1}{\pi}\left (\frac{3}{2} t(G) -\sum_i t(r_i) \left (\frac{N_i}{2} -\sum_a \gamma^{(a,i)}\right)\right )\l_G \, .
 \eea
This coincides with the numerator of the NSVZ $\b$-function \cite{Novikov:1983uc,Shifman:1986zi}
and vanishes at the fixed point.
The connection between space-time dependent couplings and the denominator of NSVZ 
has been studied in \cite{Babington:2005vu}.

\section{The super-Weyl anomaly and the Zamolodchikov metric}
\label{app_metric}

In this appendix we explain the relation between the Zamolodchikov metric and the super-Weyl anomaly coefficient $g_{I\bJ}$ defined in \eqref{eq_g_IJ}
\bea
\AA_\s^\sW \supset \int d^8z E^{-1}\sigma g_{I\bJ} \N^2\l^I \bN^2\bl^I~.
\eea
The following is a variation of the argument appearing in \cite{Osborn:1998qu}, which discusses the $c$ anomaly coefficient.

The Zamolodchikov metric $\mathcal {G}_{I\bJ}$ is defined as the matrix controlling the two point function of marginal operators at the fixed point. It is convenient to use the following expression
\bea
\langle O_I(z_1) \bar O_\bJ(z_2)\rangle
&=&\mathcal {G}_{I\bJ}
\bar {\mathcal D}^2 \mathcal D^2\Box  \left (\frac{ 1}{(x_1-x_2)^{6}}\delta^{(4)}(\theta_1-\theta_2)\right)~.
\eea
Notice that the singularity at $x_1=x_2$ has to be regulated. Using the method of differential regularization\cite{Osborn:1998qu}, we replace $\frac{1}{x^6}$ with 
\bea
\mathcal R\left (\frac{1}{x^6}\right)
&=&
-\frac{1}{32} \Box^2 \left (\frac 1{x^2} \ln (\mu^2 x^2)\right)
\eea
where $\mu$ is an arbitrary renormalization scale. The dependence of the regulated correlator on the scale $\mu$ is thus given by 
\bea
\label{eq_mu_dmu_OI_OJ}
\mu\frac{\d}{\d \mu} \langle O_I(z_{1}) \bar O_\bJ(z_{2})\rangle
&=&
\frac {\pi^2} 4  \mathcal G_{I\bJ} \bar {\mathcal D}^2 \mathcal D^2\Box  \delta^{(8)}(z_1-z_2)
\eea

On the other hand, the dependence on $\mu$ can be found by noticing that the generating functional $\WW$ is invariant under a change of renormalization scale combined with a global rescaling of the chiral compensator
\bea
\mu \frac{\d}{\d \mu}   \WW &=&\left ( \int d^6z \varphi\frac{\delta}{\delta \varphi} + c.c.\right) \WW ~.
\eea
Using the SLRG equation at the fixed point 
we can now write 
\bea
\mu\frac{\d}{\d \mu} \langle O_I(z_{1}) \bar O_\bJ(z_{2})\rangle
&=&-\frac{\delta}{\delta \lambda^I(z_1)} \frac{\delta}{\delta \lambda^I(z_2)}\AA_{\s=1}^{SW}\Big|_{\l=0}\nl
&=&
-\frac 1 8 (g_{I\bJ}+\bar g_{\bJ I}) \bar {\mathcal D}^2 \mathcal D^2\Box
 \delta^{(8)}(z_1-z_2)~.
\eea
Comparing with \eqref{eq_mu_dmu_OI_OJ} we conclude that 
\bea
  \mathcal G_{I\bJ}\propto g_{I\bJ}+\bar g_{\bJ I} ~.
\eea


\end{document}